%% file: main.tex
\documentclass[journal]{IEEEtran}
\usepackage{psfrag}
\usepackage{amsmath}
\usepackage{amssymb}
\usepackage{color}
\usepackage{graphicx}
\usepackage{epsfig}
\usepackage{fancybox}	
\usepackage{subfigure}
\usepackage{cite}
\usepackage{algorithm}
\usepackage{algorithmic}
\usepackage{multirow}
\usepackage{theorem}
\usepackage{url}
\pdfoutput=1
\usepackage{algorithm}
\usepackage{algorithmic}
\usepackage{color}	
\usepackage{siunitx}
\usepackage{multicol}	
\usepackage{mathtools}
\usepackage{epstopdf} 
\usepackage{setspace}
\usepackage{array}
\usepackage{siunitx}
\usepackage{soul}

\graphicspath{ {Figures/} }
\DeclareMathSizes{9}{9}{9}{9}
\newcommand{\ignore}[1]{}

\newcolumntype{L}[1]{>{\raggedright\let\newline\\\arraybackslash\hspace{0pt}}m{#1}}
\newcolumntype{C}[1]{>{\centering\let\newline\\\arraybackslash\hspace{0pt}}m{#1}}
\newcolumntype{R}[1]{>{\raggedleft\let\newline\\\arraybackslash\hspace{0pt}}m{#1}}

\begin{document}		
\title{CoMET: Composite--Input Magnetoelectric--based Logic Technology}
\author{Meghna G. Mankalale,~\IEEEmembership{Student Member, IEEE},
Zhaoxin Liang, Zhengyang Zhao, \\Chris H.
Kim,~\IEEEmembership{Senior Member, IEEE}, Jian--Ping Wang,~\IEEEmembership{Fellow, IEEE}, and Sachin S.
Sapatnekar,~\IEEEmembership{Fellow, IEEE}
 \thanks{All the authors are with the Department of Electrical and Computer Engineering, University of
 Minnesota, Minneapolis, MN, USA (email: manka018@umn.edu). Copyright
(c) 2017 IEEE. Personal use of this material is permitted. However,
permission to use this material for any other purposes must be
obtained from the IEEE by sending a request to pubs-permissions@ieee.org }}

\maketitle

\input{sections/abstract.tex}


\input{sections/intro.tex}

\input{sections/device.tex}

\input{sections/modeling.tex}
\input{sections/results.tex}

\input{sections/conclusion.tex}

\input{sections/ack.tex}

\IEEEpeerreviewmaketitle

\bibliographystyle{IEEEtran}
\bibliography{ref.bib}

\end{document}

%% file: sections/abstract.tex
\begin{abstract}
This work proposes CoMET, a fast and energy-efficient spintronics device for
logic applications.  An input voltage is applied to a ferroelectric (FE)
material, in contact with a composite structure -- a ferromagnet (FM) with
in-plane magnetic anisotropy (IMA) placed on top of an intra-gate FM
interconnect with perpendicular magnetic anisotropy (PMA).  Through the
magnetoelectric (ME) effect, the input voltage nucleates a domain wall (DW) at
the input end of the PMA-FM interconnect. An applied current then rapidly
propagates the DW towards the output FE structure, where the inverse-ME effect
generates an output voltage.  This voltage is propagated to the input of the
next CoMET device using a novel circuit structure that enables efficient device
cascading.  The material parameters for CoMET are optimized by systematically
exploring the impact of parameter choices on device performance.  Simulations
on a 7nm CoMET device show fast, low-energy operation, with a delay/energy of
99ps/68aJ for INV and 135ps/85aJ for MAJ3.
\end{abstract}

%% file: sections/intro.tex
\section{Introduction}

Several spin-based devices have been proposed as alternatives to
CMOS~\cite{BeyondCMOS,BeyondCMOS2015}, leveraging spin-transfer torque
(STT)~\cite{behin2010proposal,STMG,STTDW,stt-mtj}, switching a ferromagnet (FM) by
transferring electron angular momentum to the magnetic moment; spin-Hall effect
(SHE), generating spin current from a charge current through a high resistivity
material~\cite{CSL}; magnetoelectric (ME) effect~\cite{Fiebig2005}, using an electric field to
change FM magnetization~\cite{IntelME,meso}; domain wall (DW) motion through an
FM using automotion~\cite{IntelME,NikonovAutomotion}, an external field~\cite{AllwoodDW} or 
current~\cite{MTJLogic,MegDRC2016,STTDW}; dipole coupling between 
the magnets~\cite{NML}; and propagating spin wave through an
FM~\cite{swdKhitun}. In order for the spin-based processor to be running
at a CMOS--competitive clock speed of 1GHz, we need the device delay to be
around \SI{100}{ps}. Theoretically, some of the proposed devices can achieve
this target delay~\cite{BeyondCMOS2015} at the cost of additional energy.
However, in order to be competitive with CMOS, spin--based device not
only has to be fast, but also energy efficient, i.e., its 
energy dissipation should be in the range of a few hundred aJ.

We propose CoMET, a novel device that nucleates a DW in an FM channel with
perpendicular magnetic anisotropy (PMA), and uses current-driven DW motion to
propagate the signal to the output. A voltage applied on an input ferroelectric (FE) 
capacitor nucleates the DW through the ME effect. For fast, energy-efficient nucleation, 
we use a composite structure with an IMA--FM layer above the PMA--FM channel. 
The DW is propagated to the output end of the PMA channel using a charge 
current applied to a layer of high resistivity material placed under the PMA channel. 
The inverse--ME (IME) effect induces a voltage at the output end, and we use a
novel circuit structure to transmit the signal to the next stage of
logic. 

The contributions of our work are as follows:
\begin{itemize}
\item
The composite structure of IMA--FM/PMA--FM allows DW nucleation under a
low applied voltage of \SI{110}{mV}. Before the application of a voltage, 
the magnetization in the PMA--FM is moved away from its easy axis 
by the strong exchange coupling between IMA--FM/PMA--FM, 
thus enabling a fast low-power DW nucleation.
\item
We use charge current to realize fast DW propagation through the PMA--FM
interconnect. The current-driven DW motion scheme has been
experimentally shown to be fast~\cite{emori2013current,yang2015domain},
with demonstrated velocities up to \SI{750}{m/s}. We choose a PMA channel for
DW motion, as against one with
in-plane magnetic anisotropy (IMA), since it is more robust to DW pinning
and surface roughness effects~\cite{PMAPinning1, PMAPinning2}.
\item 
A novel circuit structure comprising a dual--rail inverter allows efficient
cascading of devices. This scheme improves upon a previous scheme~\cite{IntelME}
of 6:1 device ratioing and the need for repeated amplifications.
\item 
We explore the design space of the possible PMA--FM material
parameters to optimize the performance of the device. Through this
systematic design space exploration, we show that it is possible to
achieve inverter delay/energy of \SI{99}{ps}/\SI{68}{aJ}. 
\end{itemize}
\vspace{-15pt}

\setcounter{figure}{0}
\begin{figure}[ht]
\centering
\includegraphics[width=6cm]{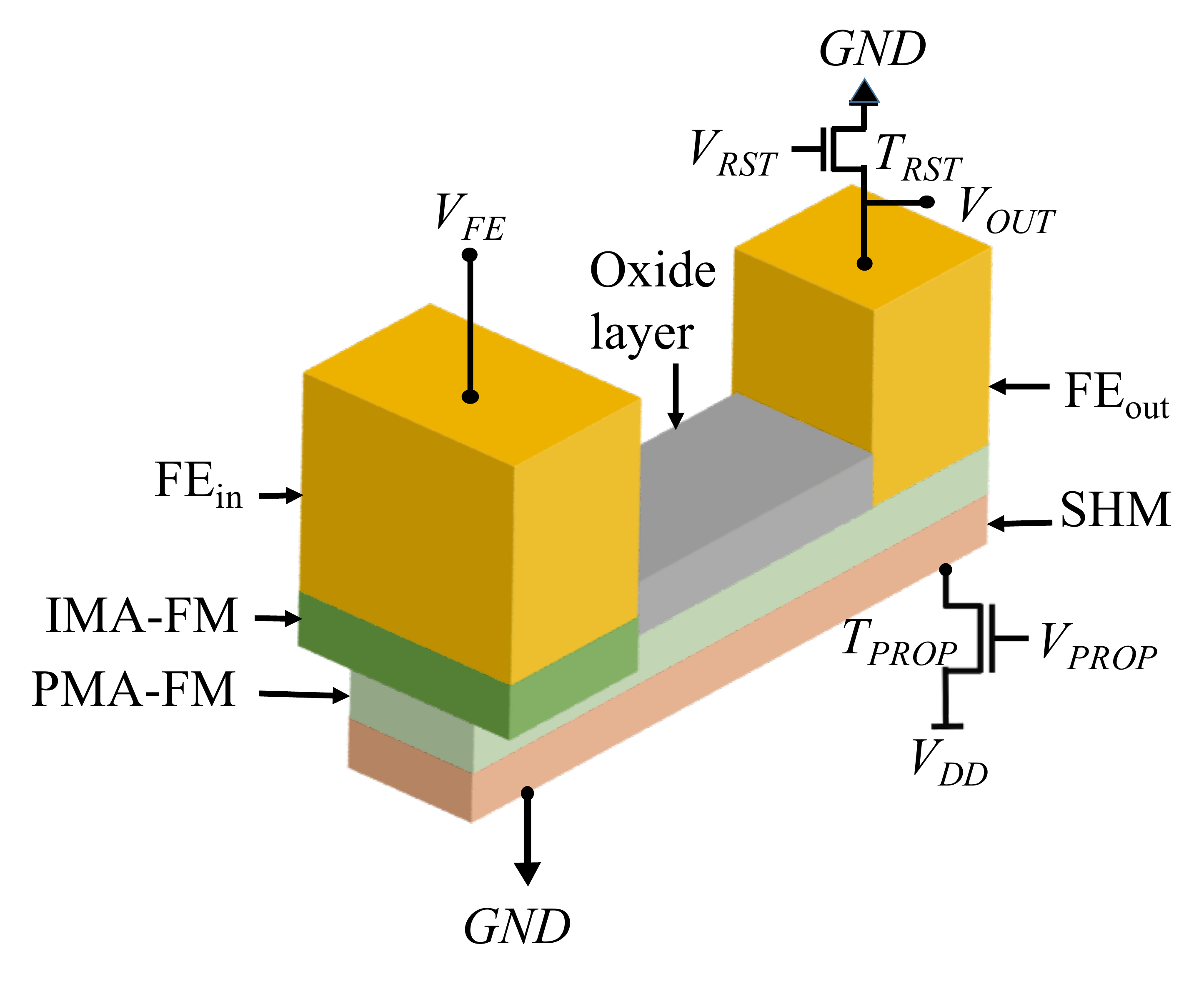}
\caption{Proposed device concept of CoMET illustrating the composite structure of IMA--FM exchange--coupled 
with PMA--FM at the input end.}
\vspace{-8pt}
\label{fig:device}
\end{figure}

The rest of this paper is organized as follows: In Section~\ref{sec:device}, we 
explain the operation of CoMET. We present the mathematical models and the 
simulation framework used in this work in Section~\ref{sec:modeling}. Next, we show the 
performance of the device as a function of the material parameters in 
Section~\ref{sec:results}. Section~\ref{sec:conclusion} concludes the paper.

%% file: sections/device.tex
\section{CoMET: Device Concept and Operation}
\label{sec:device}
The structure of the proposed device is shown in Fig.~\ref{fig:device}. At the input, a 
FE capacitor, $\text{FE}_{\text{in}}$, is placed atop an
IMA--FM. The IMA--FM is exchange--coupled with the input end of a longer
PMA--FM
interconnect.  At its output end, a second FE capacitor,
$\text{FE}_{\text{out}}$, is placed on top of the PMA--FM interconnect. A layer
of high-resistivity spin-Hall material (SHM), which is conducive to strong
spin-orbit interaction, is placed beneath the PMA--FM. An oxide layer is present
on top of PMA--FM between $\text{FE}_{\text{in}}$ and $\text{FE}_{\text{out}}$.

\subsection{CoMET--based inverter}
\begin{figure}[hb]
\centering
\includegraphics[width=8.5cm]{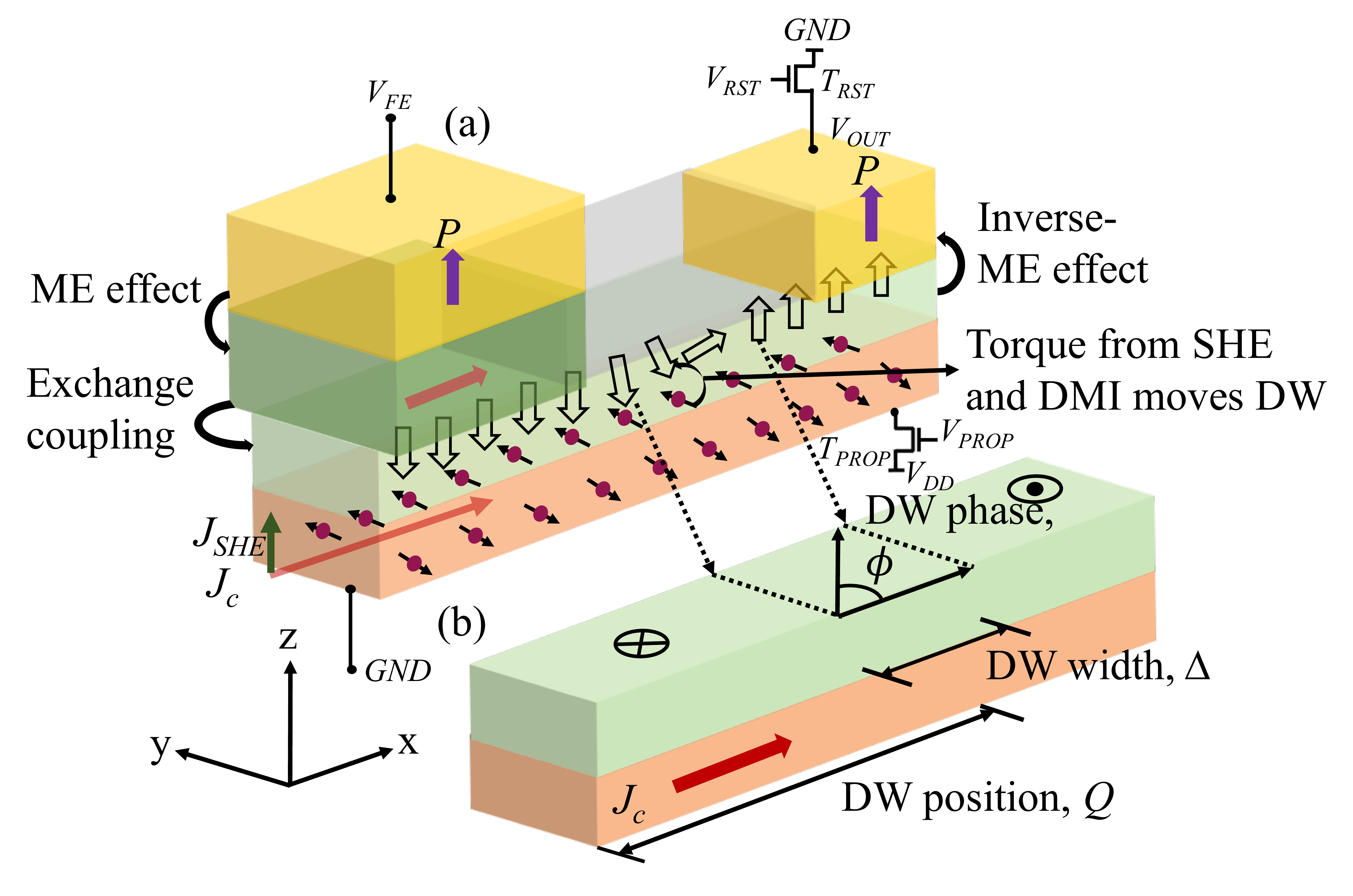}
\caption{(a) Graphical representation of the different underlying physical mechanisms of the device. (b) The 
position of the DW ({\em Q}), width ($\Delta$), and phase ($\phi$).}
\label{fig:physics}
\end{figure}

\begin{figure}[ht]
\centering
\includegraphics[width=8cm]{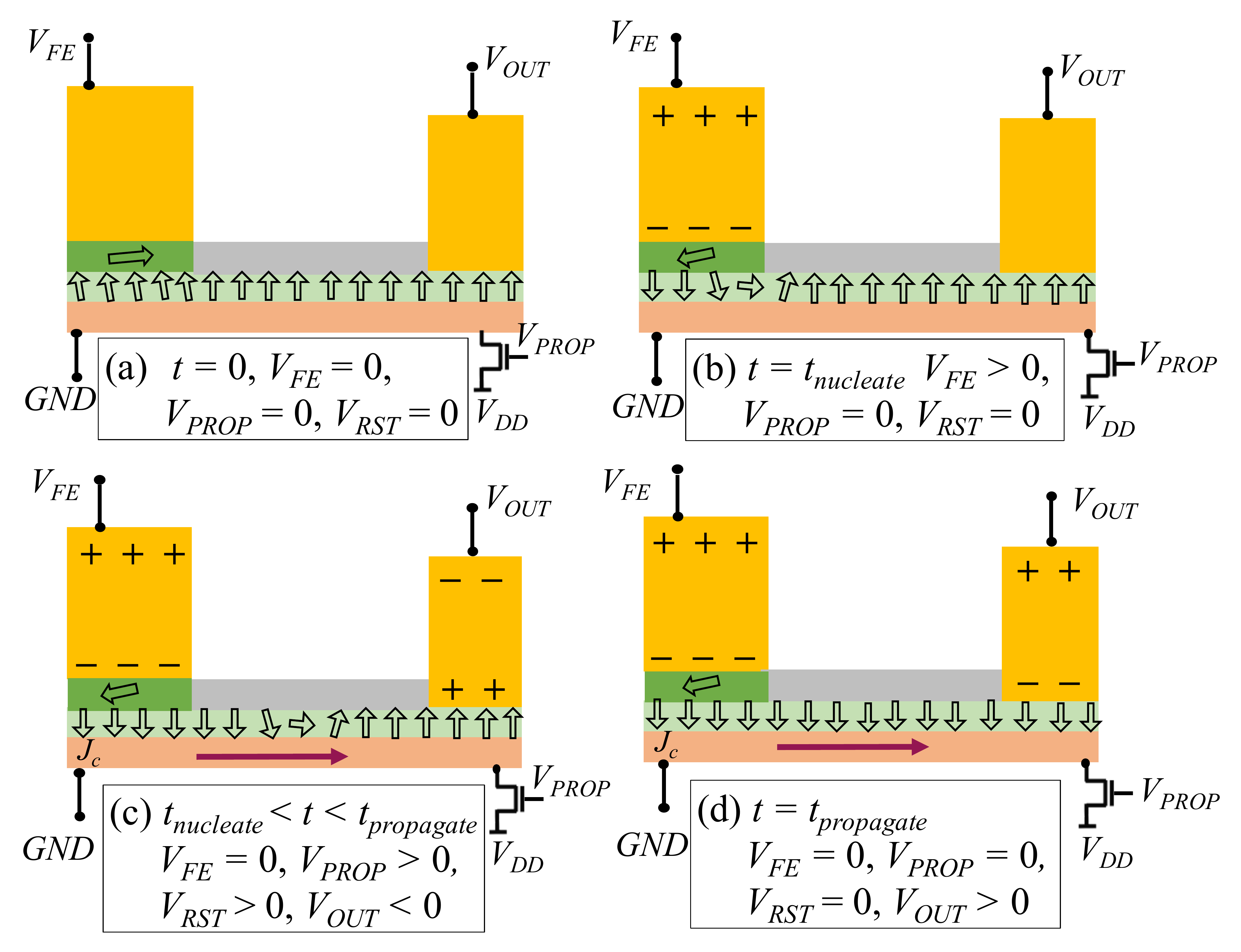}
\caption{Operation of CoMET device showing (a) steady--state before $V_{FE}$ is applied, (b) nucleation 
of the DW in PMA--FM after $V_{FE}$ is turned on, (c) propagation of the
DW by turning on $V_{PROP}$, and 
charging of the output FE capacitor when $V_{RST}>0$ is applied, and (d) the induction of an output voltage 
$V_{OUT}$ through the inverse--ME effect.}
\vspace{-8pt}
\label{fig:tdiag}
\end{figure}

We explain the device operation in four stages with the help of Fig.~\ref{fig:physics} 
and Fig.~\ref{fig:tdiag}.

\vspace{0.25cm}
\noindent
{\bf Stage 1 -- DW nucleation:}
At time $t = 0$, an applied voltage, $V_{FE}$, charges
$\text{FE}_{\text{in}}$.  The resulting electric field across
$\text{FE}_{\text{in}}$, $E_{FE}$, may be positive or negative, depending on the
sign of $V_{FE}$, and generates an effective magnetic field, $H_{ME}$, through
the ME effect that couples the electric polarization in FE$_{\text{in}}$
with the magnetization in the IMA--FM. This magnetic field acts on the composite structure.  For $V_{FE} > 0$, this
nucleates a DW in the PMA--FM as seen from Fig.~\ref{fig:tdiag}(b), 
with a down--up configuration if the initial magnetization is along the $+z$ axis. 
For the opposite case, an up--down configuration is nucleated.

If the initial orientation of the PMA--FM is at an angle to the z-axis, a
smaller $H_{ME}$ field can nucleate the DW. The composite structure creates
this angle due to strong exchange coupling between the IMA--FM and the
PMA--FM as can be seen from the magnetization of PMA--FM in
Fig.~\ref{fig:tdiag}(a), thus allowing nucleation under a low magnitude
of $V_{FE}$. In the absence of IMA--FM, voltages up to \SI{1}{V} are necessary
to nucleate a DW whereas we show that with the presence of
IMA--FM, voltages as low as \SI{110}{mV} would suffice.  

\vspace{0.25cm}
\noindent
{\bf Stage 2 -- DW propagation}
Once the DW is nucleated in PMA--FM, transistor $T_{PROP}$ is turned on using the signal 
$V_{PROP}$ to send a charge current ($J_c$) through the SHM. 
Due to SHE, electrons with opposite spin accumulate in the direction transverse to the charge 
current as shown in Fig.~\ref{fig:physics}. As a result, a spin current ($J_{SHE}$) is 
generated in a direction normal to the plane of SHM. The resultant torque 
from the combination of SHE and Dzyaloshinskii--Moriya interaction (DMI)~\cite{emori2013current} 
at the interface of PMA--FM and the SHM propagates the DW to the output end. 

Before the DW reaches the output, $V_{RST}$ turns on 
transistor $T_{RST}$ to connect $\text{FE}_{\text{out}}$ to {\em GND} as seen
from Fig.~\ref{fig:tdiag}(c). This causes $\text{FE}_{\text{out}}$ to
charge due to the presence of an electric field across it as a result of the IME effect.
This step resets the capacitor such that once the DW reaches
the output, it can either reverse or maintain the electric polarization
of FE$_{\text{out}}$, thus reflecting the result of the operation.

\vspace{0.25cm}
\noindent
{\bf Stage 3 -- Output FE switching:}
The DW reaches the output end in time $t_{propagate}$ as seen from
Fig.~\ref{fig:tdiag}(d) and switches the magnetization of PMA--FM. The
magnetization in PMA--FM couples with the 
electric polarization of $\text{FE}_{\text{out}}$ through the IME effect. 
As a result, a voltage, $V_{OUT}$, is induced at the output node. 

\vspace{0.25cm}
\noindent
{\bf Stage 4 -- Cascading multiple logic stages:}
Successive logic stages of CoMET can be cascaded as shown in
Fig.~\ref{fig:cascade}(a),
through a dual-rail inverter structure comprising transistors $T_P$ 
and $T_N$. A timing diagram showing the application of the different
input excitations and the output signal are shown in {Fig.~\ref{fig:cascade}}(b). 
The signal $V_{RST}$ turns on transistors $T_{RST1}$ and $T_{RST2}$ in 
the two logic stages to charge the respective FE capacitors. The output 
voltage induced through the IME effect, $V_{OUT1}$, turns on either $T_N$ or $T_P$,
depending on its polarity. These transistors form an inverter and set
$V_{FE}$ for the next stage to a polarity opposite that of $V_{OUT1}$. 
The result of the operation is retained in the PMA-FM when the supply
voltage is removed. This allows the realization of nonvolatile logic
with CoMET. As a result, the inverter can be power-gated after signal
transfer, saving leakage. Unlike the charge transfer scheme in~\cite{IntelME} 
with 6:1 ratioing between stages and repeated amplification, our scheme allows 
all stages to be unit-sized, resulting in area and energy efficiency. 
This scheme also allows efficient charge-based cascading of logic
stages as opposed to spin--based cascading, which require a large
number of buffers to overcome the spin losses in the
interconnects~\cite{Kim2015}.

\setcounter{figure}{3}
\begin{figure}[h]
    \centering
	\subfigure[]{
    	\includegraphics[width=9cm]{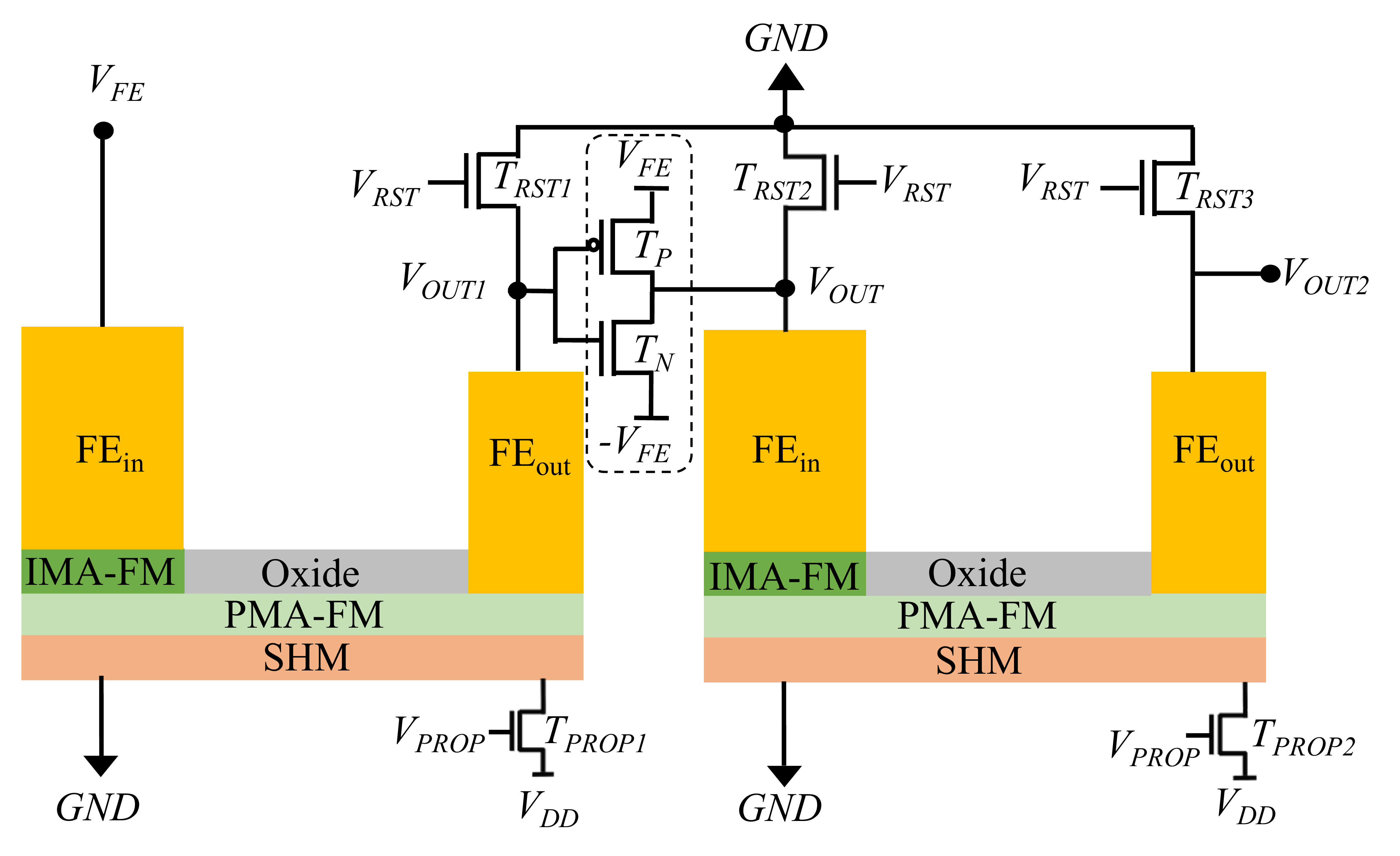}
	}

	\subfigure[]{
    	\includegraphics[width=6.5cm]{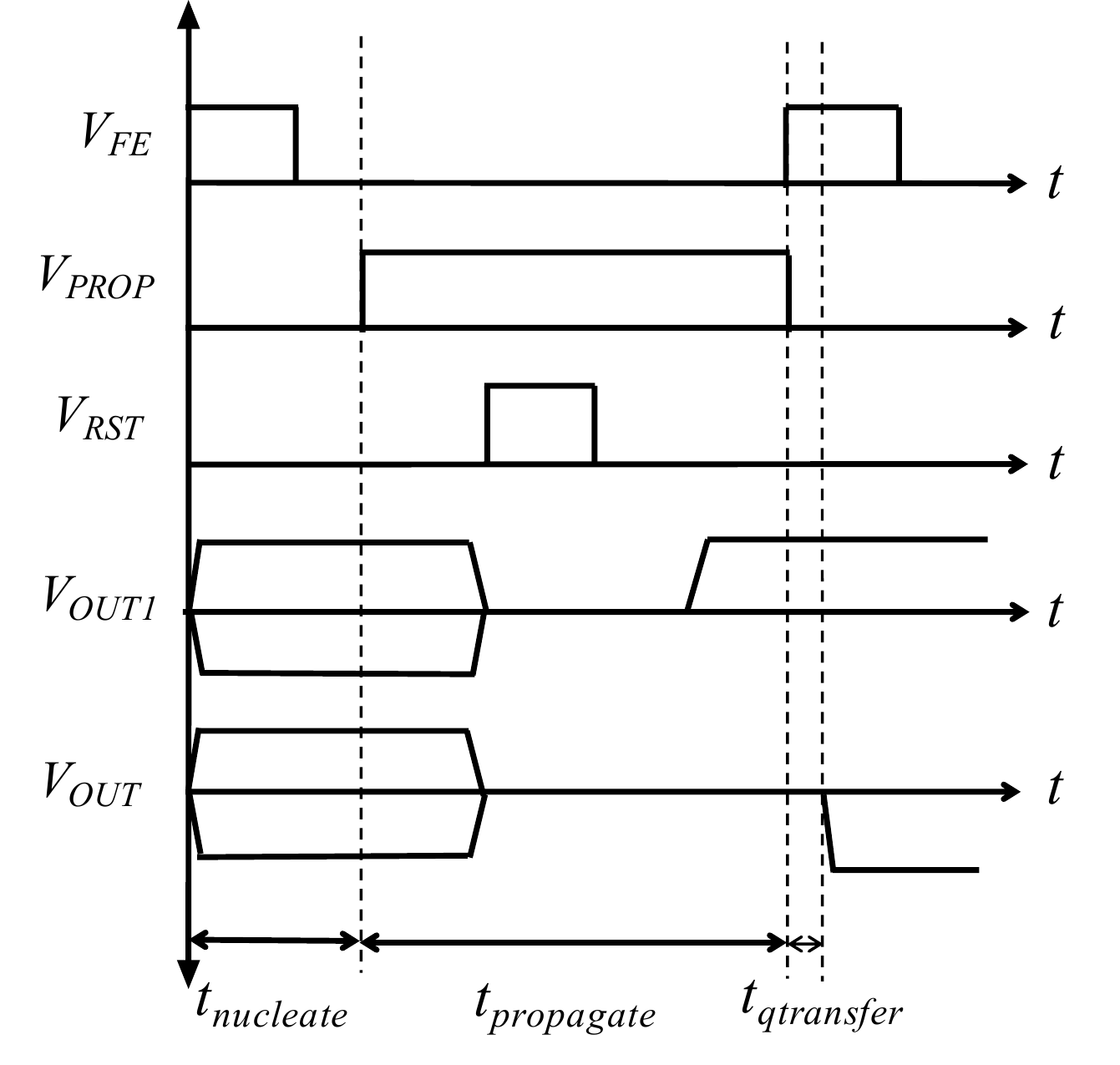}
	}
    \caption{(a) Logic cascading of two CoMET devices using transistors
$T_{P}$ and $T_{N}$, and (b) timing diagram showing the application of
the $V_{FE}$, $V_{PROP}$, and $V_{RST}$ signals.}
	\vspace{-15pt}
    \label{fig:cascade}
\end{figure}

\subsection{CoMET--based Majority gate}
\begin{figure}[h]
    \centering
	\subfigure[]{
		\includegraphics[width=7.5cm]{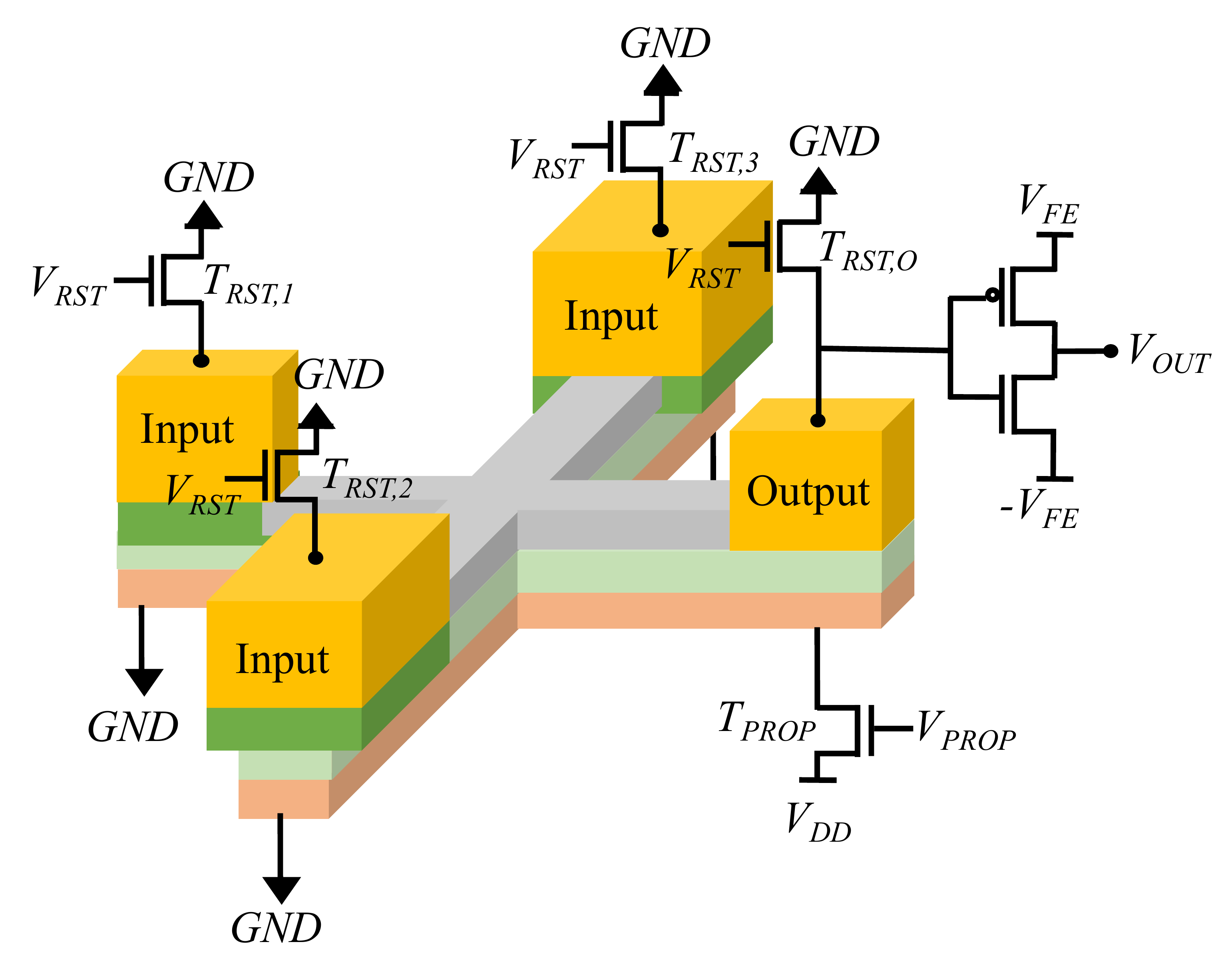}
	}

	\subfigure[] {
		\includegraphics[width=4cm]{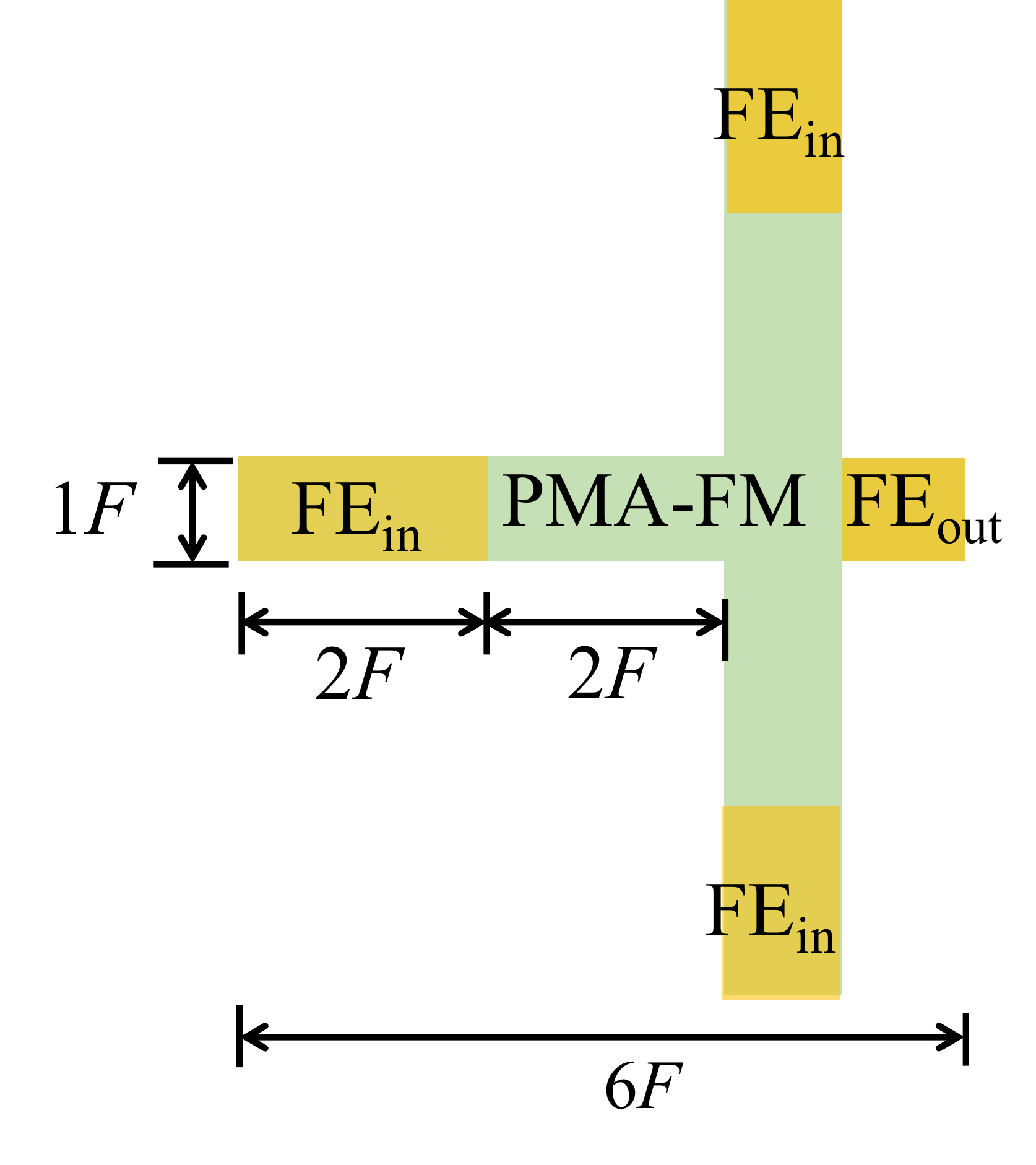}
	}
	\subfigure[] {
		\includegraphics[width=3.9cm]{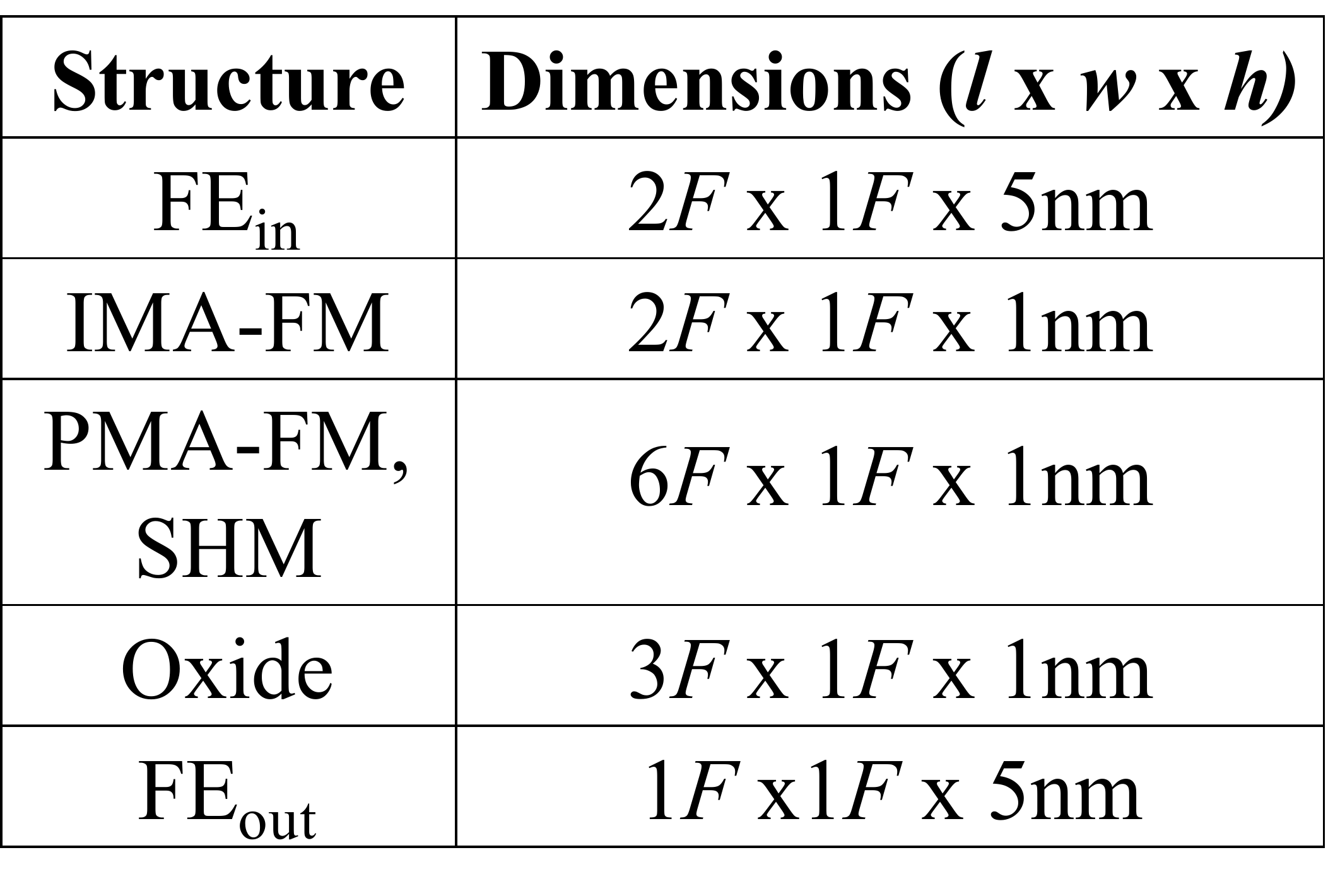}
	}
    \caption{(a) CoMET--based three-input majority (MAJ3) gate (b) top view of MAJ3 with the device 
    dimensions marked for a feature size, $F$ and (c) the length
($l$), width ($w$), and height ($h$) of the CoMET device in
Fig.~\ref{fig:device} considered in this simulation.}
    \vspace{-15pt}
	\label{fig:majority}
\end{figure}

The idea of the CoMET inverter can be extended to build a 
three-input CoMET majority gate (MAJ3), as shown in
Fig.~\ref{fig:majority}(a). The input voltage $V_{FE}$ is applied
to each input to nucleate a DW in the PMA--FM below each
$\text{FE}_{\text{in}}$.  The DWs from each input is propagated to the output by turning on
$T_{PROP}$. The DWs compete in the PMA--FM~\cite{STMG}, and the majority
prevails to switch $\text{FE}_{\text{out}}$ using the IME effect. Subsequent gates are cascaded
using the dual-rail inverter scheme described above.

%% file: sections/modeling.tex
\section{Modeling and simulation framework}
\label{sec:modeling}
We now show how the performance of a MAJ3 gate can be modeled.
The worst--case delay of this gate occurs when one input differs from the
others.  At feature size, $F$, the DW for each input nucleates in
PMA--FM below FE$_{\text{in}}$ at a distance $2F$ once $V_{FE}$ is
applied. The DW from each input then travels a
$4F$ distance to switch $\text{FE}_{\text{out}}$ as shown in Fig.~\ref{fig:majority}(b).  
The dimensions of the simulated structure are shown in {Fig.~\ref{fig:majority}(c)}. The IMA--FM
aspect ratio (x:y) is set to 2:1 to align the magnetization of PMA--FM
at an angle to the easy axis (due to shape
anisotropy).  The $\text{FE}_{\text{in}}$
and $\text{FE}_{\text{out}}$ thicknesses are set to \SI{5}{nm} to avoid
leakage through the capacitors. The PMA--FM thickness is set to \SI{1}{nm}. 

\subsection{Modeling device operation}
We analyze the device operation in each of the four stages as follows:

\vspace{0.25cm}
\noindent
{\bf Stage 1 -- DW nucleation:} 
The dynamics of electric polarization, $\vec{P}$, of $\text{FE}_{\text{in}}$ 
due to $E_{FE}(= V_{FE}/h_{\text{{FE}}_{\text{in}}})$ as a result of the applied voltage
$V_{FE}$ across the thickness of the input FE capacitor,
$h_{{\text{FE}}_{\text{in}}}$
are described by the Landau-Khalatnikov (LKh) equation~\cite{LKhmodel}: 
\begin{equation}
\gamma_v\frac{\partial {P_i}}{\partial t} =
-\frac{1}{a_{{\text{FE}_\text{in}}}}\frac{\partial
{F}_T}{\partial {P_i}}
\label{eq:LKh}
\end{equation}

\noindent
where ${F}_T$ is the total free energy of the input structure 
as a function of $E_{FE}$, $\gamma_v$ is the viscosity coefficient,
$P_i$ is the component of $\vec{P}$ in the $i$ direction, and 
$a_{{\text{FE}_{\text{in}}}}$ is the volume of the input FE capacitor.  
The resultant $\vec{P}$ generates an effective magnetic field from ME,
$\vec{H}_{ME}$ given by,

\begin{equation}
    \vec{H}_{ME} = \frac{\kappa_{ME}}{\epsilon_0}\frac{h_{int}}{h_{{\text{FE}}_{\text{in}}}}\vec{P}
\end{equation}

\noindent
Here, $h_{int}$ is the ME interface thickness, $h_{\text{FE}_{\text{in}}}$ denotes the 
thickness of FE$_{\text{in}}$, and $\kappa_{ME}$ refers to the ME
coefficient. The magnetic field, $H_{ME}$ is then 
applied as a Zeeman field to the composite structure in the micromagnetics 
simulator, OOMMF~\cite{OOMMF}, which solves the Landau-Lifshitz-Gilbert
(LLG) equation~\cite{landaulifshitz,gilbert} as shown below, 
to obtain $t_{nucleate}$:

\begin{equation}
    \frac{(1+\alpha^2)}{\gamma}\frac{d\vec{M}}{dt} =
-\vec{M}{\times}\vec{H}_{eff} - \alpha\vec{M}{\times}(\vec{M}{\times}\vec{H}_{eff})
    \label{eq:LLG}
\end{equation}

\noindent
Here $\alpha$ refers to the damping constant and $\vec{M}$ denotes the magnetization in PMA--FM.  
The effective magnetic field, $\vec{H}_{eff}$ is given by:

\begin{equation}
    \vec{H}_{eff} = \vec{H}_{ME} + \vec{H}_{K} + \vec{H}_{demag} +
\vec{H}_{ex} 
\end{equation}

\noindent 
where $\vec{H}_{K}$, $\vec{H}_{demag}$, and $\vec{H}_{ex}$ refer to the 
contributions to $\vec{H}_{eff}$ from magnetic anisotropy, 
the demagnetization field, and the exchange field in PMA--FM,
respectively.

\vspace{0.25cm}
\noindent
{\bf Stage 2 -- DW propagation:}
The 1D equations that model DW motion describe its instantaneous
velocity, $dQ/dt$ and phase $\phi$~\cite{emori2013current,thiavilleDMI} 
(defined in Fig.~\ref{fig:physics}) through a pair of coupled
differential equations:
  
\begin{equation}
    \begin{aligned}
        (1+\alpha^2) \frac{dQ}{dt} & \ = \ -\gamma \Delta \frac{H_K}{2} \sin(2\phi) + (1+\alpha^2\beta)B_{STT} \\ 
                                   & + \gamma \Delta \frac{\pi}{2} [\alpha H_{SHE}cos(\phi) + H_{DMI}\sin(\phi)] \\
        (1+\alpha^2) \frac{d\phi}{dt} & \ = \ -\gamma\alpha \frac{H_K}{2}\sin(2\phi) + \frac{(\beta - \alpha)}{\Delta}B_{STT} 
        \\ & + \gamma\frac{\pi}{2}[H_{SHE}\cos(\phi) + \alpha   H_{DMI}\sin(\phi)]
    \end{aligned}
\end{equation}
\noindent
The DW width, $\Delta$~\cite{thomas2007current} is given by, 
\begin{equation}
    \Delta = \frac{\sqrt{A/K_{\text{U,PMA--FM}}}}{\sqrt{1+\frac{\mu_0{M_{\text{S,PMA--FM}}}^2}{K_{\text{U,PMA--FM}}}[\frac{h_{\text{PMA--FM}}}{h_{\text{PMA--FM}}+\Delta} - \frac{h_{\text{PMA--FM}}}{h_{\text{PMA--FM}} + w_{\text{PMA--FM}}}]\sin^2(\phi)}}
\end{equation}
\noindent
whereas the effective field from anisotropy ($H_K$), SHE ($H_{SHE}$), 
DMI ($H_{DMI}$), and field-like term from STT ($B_{STT}$) is given by, 
\begin{equation}
	\begin{aligned}
		H_K   &= \frac{2K_{\text{U,PMA--FM}}}{{M_{\text{S,PMA--FM}}}};
	  H_{SHE} = \frac{{\hbar}{\theta_{SHE}}J_c}{2\mu_0eM_{\text{S,PMA--FM}}} \\	
	  H_{DMI} &= \frac{D}{\mu_0M_{\text{S,PMA--FM}}\Delta};
	B_{STT}   = \frac{{\mu_B}P_{\text{PMA--FM}}J_c}{eM_{\text{S,PMA--FM}}}
	\end{aligned}
\label{eq:fields}
\end{equation}

\ignore{
\noindent
 The effective field from anisotropy ($H_K=~(2K_{\text{U,PMA--FM}})/({M_{\text{S,PMA--FM}}})$), SHE
($H_{SHE}=~({\hbar}{\theta_{SHE}}J_c)/({2\mu_0eM_{\text{S,PMA--FM}}})$), 
and DMI ($H_{DMI}=~D/(\mu_0M_{\text{S,PMA--FM}}\Delta)$) propagate the DW
through the interconnect. The effective field-like term from STT,
($B_{STT} = {\mu_B}P_{\text{PMA--FM}}J_C/{eM_{\text{S,PMA--FM}}}$)
has little effect on the motion of the DW through the
PMA--FM~\cite{emori2013current}.}

\noindent
The contribution of $B_{STT}$ to the motion of the DW in PMA--FM is
negligible compared to those from SHE and DMI~\cite{emori2013current}.
Here, $A$, $M_{\text{S,PMA--FM}}$, $P_{\text{PMA--FM}}$, $h_{\text{PMA--FM}}$, 
$K_{{\text{U,PMA--FM}}}$, $\beta$, $\theta_{SHE}$, and $D$ refer to the exchange constant, 
PMA--FM saturation magnetization, PMA--FM polarization ratio, PMA--FM
thickness, PMA--FM uniaxial anisotropy, adiabatic STT parameter, spin-Hall angle, and DMI constant, respectively. 
The average DW velocity is used to calculate $t_{propagate}$.  

\begin{table}[h]
     \centering
     \label{tbl:parameters}
     \begin{tabular}{|L{6.4cm}|L{2cm}|}
         \hline
         {\bf Parameter}                                                                          & {\bf Value}                           \\ \hline
         Viscosity coefficient, ${\gamma}_v$ {[}Vm$\cdot$s/K{]}~\cite{IntelME}                    & 5.47$\times$10$^{-5}$                 \\ \hline
         Vacuum permittivity, $\epsilon_0$ {[}F/m{]}                                              & 8.85$\times$10$^{-12}$                \\ \hline
         Vacuum permeability, $\mu_0$ {[}T$\cdot$m/A{]}                                           & 1.25$\times$10$^{-6}$                 \\ \hline
         Charge of the electron, $e$ [C]                                                          & 1.60$\times$10$^{-19}$	              \\ \hline
		 Gyromagnetic ratio, $\gamma$ {[}rad/s$\cdot$T{]}                                         & 1.76$\times$10$^{11}$                 \\ \hline
		 Speed of light, c {[}m/s{]}                                                              & 3$\times$10$^8$                       \\ \hline
         ME coefficient for FE$_\text{in}$, $\kappa_{ME}$ {[}s/m{]}~\cite{benchmarkingME}         & (0.2/c)                     		  \\ \hline
         ME coefficient for FE$_\text{out}$, $\kappa_{IME}$ {[}s/m{]}~\cite{benchmarkingME}       & (1.4/c)                      		  \\ \hline
         Resistivity of SHM, $\rho_{\text{SHM}}$ {[}$\Omega$-m{]}~\cite{emori2013current}         & 1.06$\times$10$^{-7}$                 \\ \hline
         FE permittivity, $\epsilon_{FE}$~\cite{IntelME}                                          & 164                                   \\ \hline
         Adiabatic STT parameter, $\beta$~\cite{emori2013current}                                 & 0.4                                   \\ \hline
         DMI constant, $\lvert{D}\rvert$ {[}mJ/m$^2${]}~\cite{emori2013current, thiavilleDMI}     & 0.8                                   \\ \hline
         ME interface thickness, $h_{int}$ {[}nm{]}~\cite{IntelME}                                & 1.5                                   \\ \hline
         Transistor threshold voltage, $V_{th}$ {[}V{]}~\cite{PTM}                                & 0.2                                   \\ \hline
         Bohr magneton, $\mu_B$ {[}J/T{]}                                                         & 9.274$\times$10$^{-24}$               \\ \hline
         $\SI{15}{nm}$ Transistor on-resistance, $R_{on}$ {[}$\Omega${]}~\cite{PTM}                & 3480                                  \\ \hline
         $\SI{7}{nm}$ Transistor on-resistance, $R_{on}$ {[}$\Omega${]}~\cite{PTM}                 & 4109                                  \\ \hline
         Spin Hall angle, $\theta_{\text{SHE}}$                                                   & 0.5                                   \\ \hline
         Spin polarization, $P_{\text{PMA--FM}}$~\cite{emori2013current}                          & 0.5                                   \\ \hline
         Transistor gate capacitance, $C_g$ {[}fF{]}~\cite{PTM}                                   & 0.1                                   \\ \hline
     \end{tabular}
     \caption{Simulation parameters used in this work.}
	\vspace{-20pt} 
\end{table}

\vspace{0.25cm}
\noindent
{\bf Stage 3 -- Output FE switching:}
The electric field developed across $\text{FE}_{\text{out}}$ 
from IME effect, $\vec{E}_{IME}$, due to the magnetization,
$\vec{M}$ in PMA--FM is used to calculate  $V_{OUT}$ as shown below:
\begin{equation}
    \begin{aligned}
        \vec{E}_{IME} &= \kappa_{IME}\frac{h_{int}}{h_{\text{FE}_{\text{out}}}}\vec{M};\\
        V_{OUT} &= \vec{E}_{IME}h_{\text{FE}_{\text{out}}}
    \end{aligned}
\end{equation}
\noindent
where $\kappa_{IME}$ is the inverse ME
coefficient~\cite{benchmarkingME}, $h_{int}$ is the interface thickness,
and $h_{\text{FE}_{\text{out}}}$ refers to the thickness of the output
FE capacitor. 

\vspace{0.25cm}
\noindent
{\bf Stage 4 -- Cascading logic stages:}
The time, $t_{qtransfer}$, required to transfer $V_{OUT1}$ to the input of the
next stage includes the delay of the dual-rail inverter and the RC delay of the
wire from the inverter output to $\text{FE}_{\text{in}}$ of the next stage.  
 
\begin{figure}[h]
\centering
\includegraphics[width=9cm]{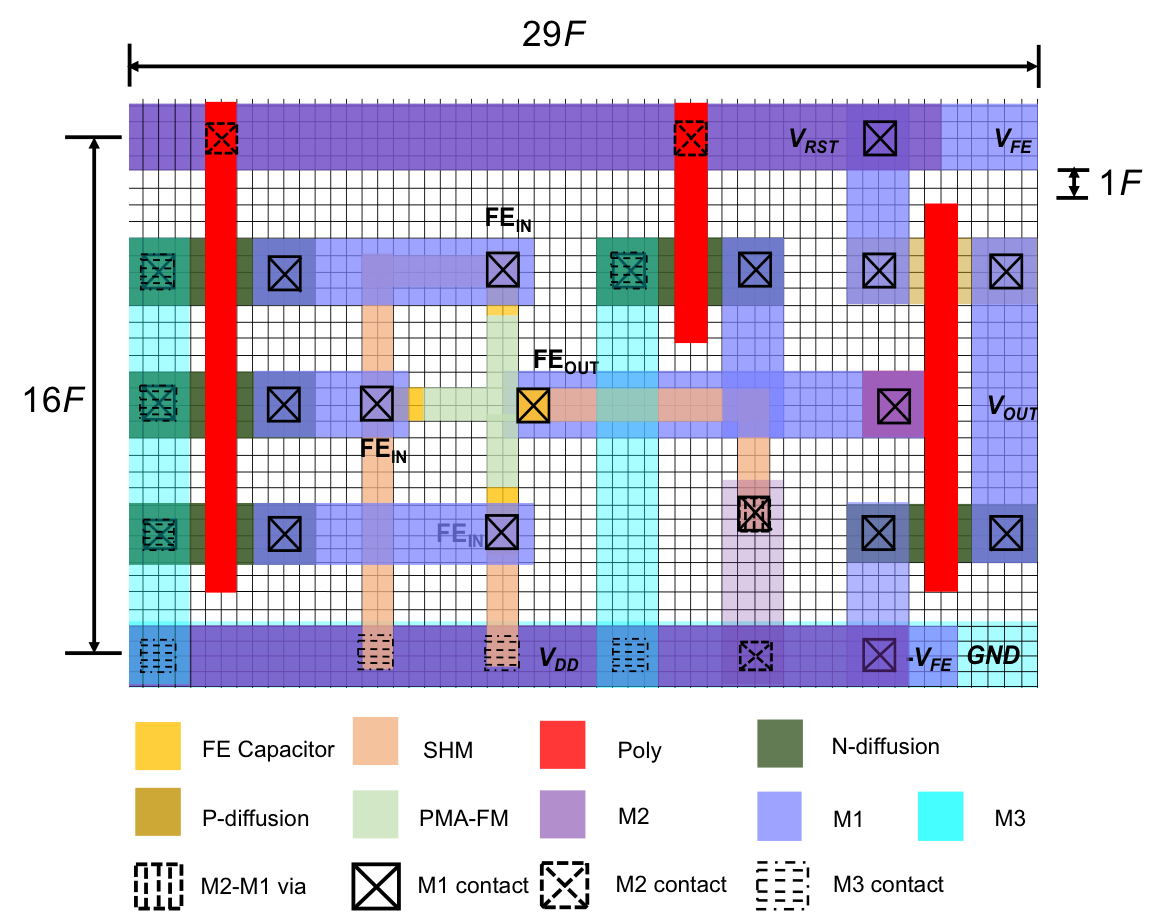}
\caption{Layout of a CoMET--based three-input majority gate.}
\vspace{-8pt}
\label{fig:layout}
\end{figure}

\begin{figure*}[ht]
      \subfigure[]{
		 \includegraphics[width=0.32\textwidth]{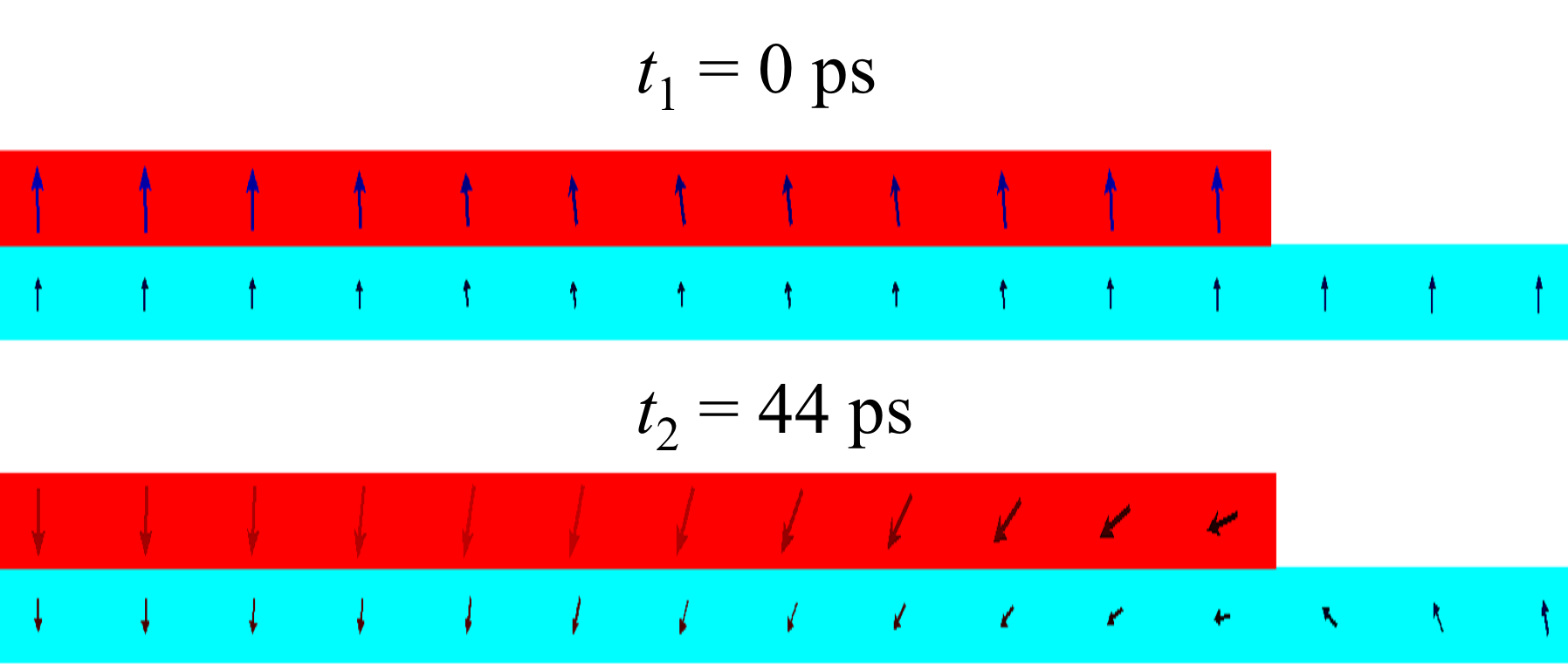}
      }
	  \subfigure[]{
		 \includegraphics[width=0.32\textwidth]{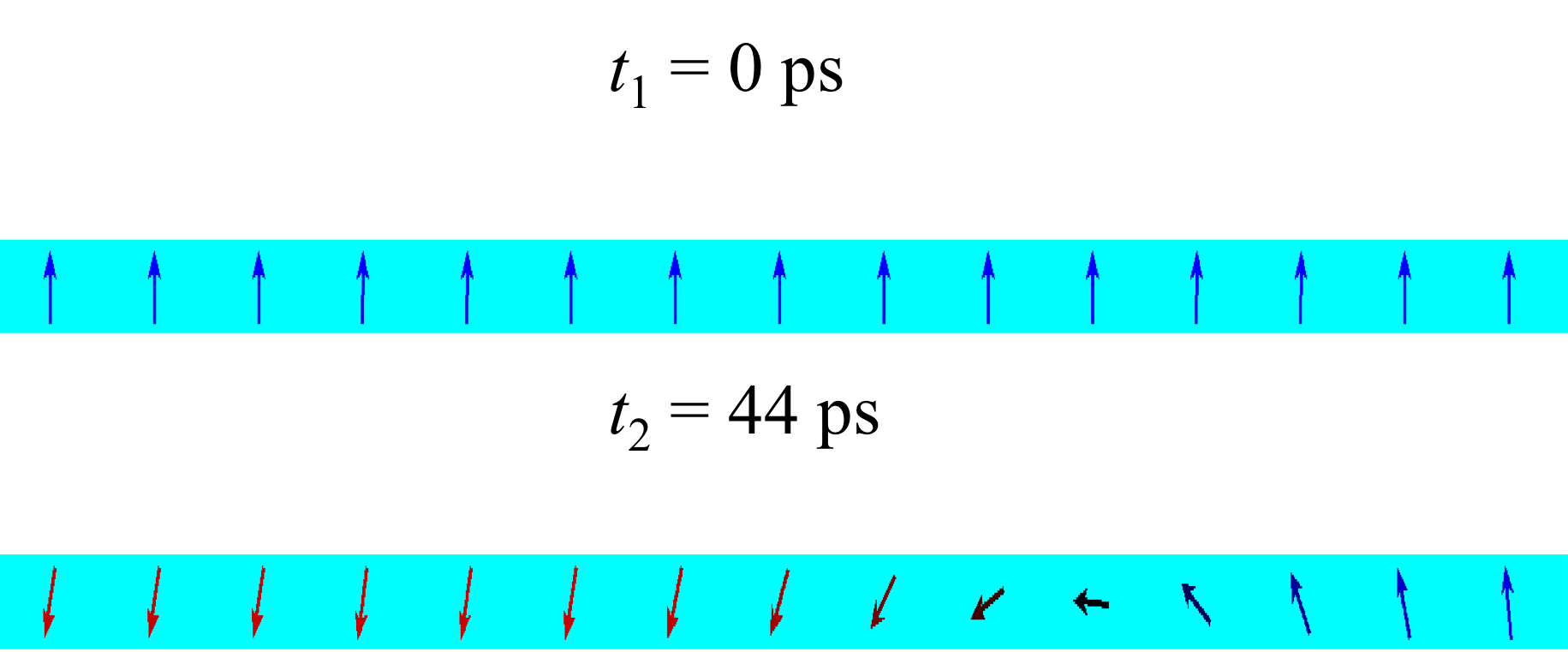}
      }
	  \subfigure[]{
		 \includegraphics[width=0.32\textwidth]{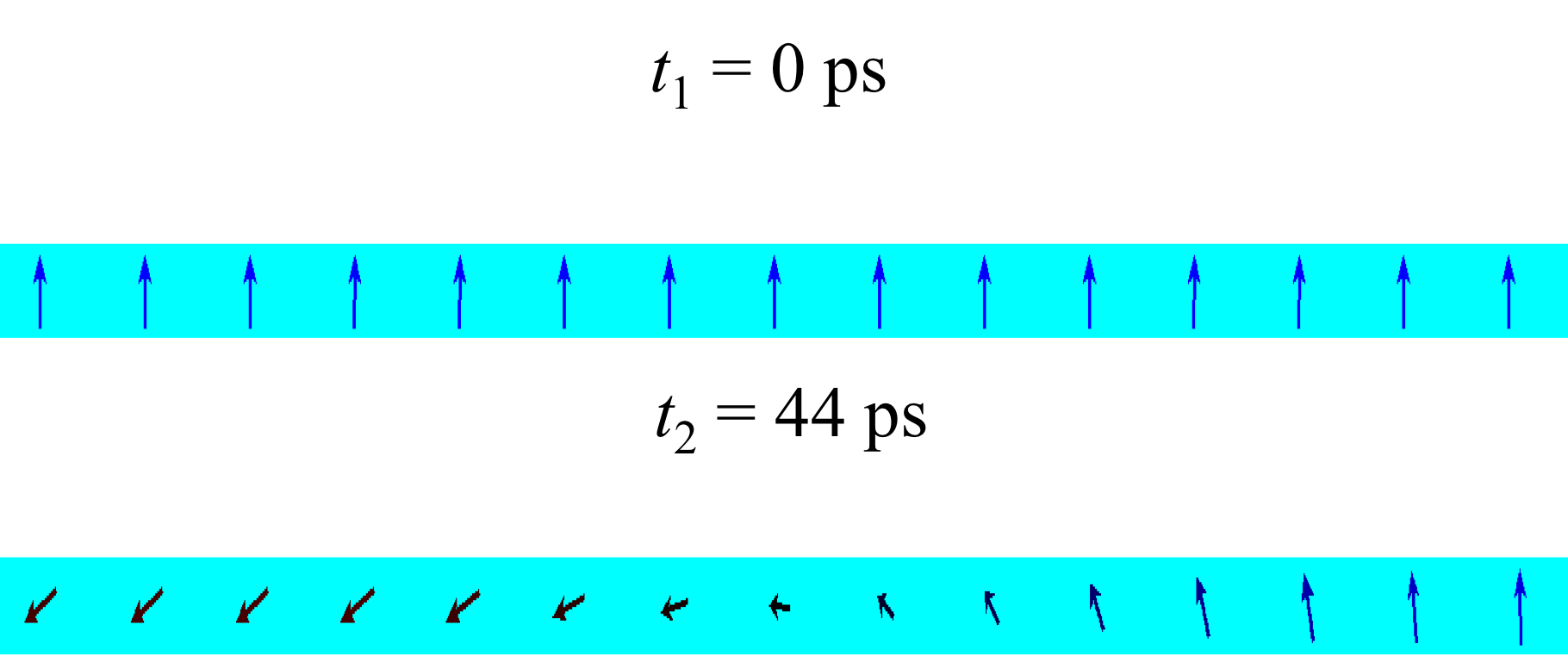}
      }
	\caption{DW nucleation in PMA--FM (a) with the composite structure used
	in this work (b) without the composite structure, i.e., without the IMA--FM with the
	ME field applied for a region $2F\times1F\times1$nm from the left end of
	PMA--FM and (c) without the composite structure with the ME field
	applied for a region $1F\times1F\times1$nm from the left end of PMA--FM.
	In (a) the red region refers to the IMA--FM, and the blue region refers
	to PMA--FM in (a), (b), and (c). 
	The material parameters used in the OOMMF simulation are:
	$M_{\text{S,PMA--FM}} = \SI{0.5e6}{A/m}$, $K_{\text{U,PMA--FM}} =
	\SI{0.6e6}{J/m^3}$, $A=\SI{10}{pJ/m}$, $\alpha=0.01$. The voltages
	required to nucleate the DW at $t_{nucleate} = \SI{44}{ps}$ corresponding to
	(a) $V_{FE}=\SI{110}{mV}$, (b) $V_{FE} = \SI{350}{mV}$ and (c) $V_{FE} =
	\SI{1.06}{V}$.}
	\vspace{-8pt}
	\label{fig:OOMMF}
\end{figure*}

\subsection{Modeling performance parameters}
The delay and energy of a $K$-input CoMET majority gate are:
\begin{equation}
\begin{aligned}
T_{\text{CoMET}} = & 2(t_{nucleate} + t_{propagate} + t_{qtransfer})\\
E_{\text{CoMET}} = & 2(E_{FE} + E_{TX} + E_{Joule} + E_{leakage}) 
\end{aligned}
\label{eq:delayenergy}
\end{equation}
where $E_{FE}$, $E_{TX}$, $E_{Joule}$, and $E_{leakage}$, respectively, refer
to the energy for charging the $\text{FE}_{\text{in}}$, turning the transistors
on, SHM Joule heating, and due to transistor leakage currents.  The factor of
$2$ is due to PMA--FM magnetization initialization of each input to a state that
allows DW nucleation~\cite{IntelME}. Finally,
\begin{align*}
	E_{TX} &= (C_g/2) \left ((K+1)V_{RST}^{2}+ V_{PROP}^2 + 2V_{OUT}^2
	\right ) ; \\
	E_{Joule} &= {(J_{c}w_{\text{SHM}}t_{\text{SHM}})}^2\big[R_{on} +
	R_{\text{SHM}}\big]t_{propagate} ; \\
	E_{FE} &= (K/2) C_{\text{FE}_{\text{in}}}V_{FE}^2 \; \; ; \; \;
	R_{\text{SHM}} =
	(\rho_{\text{SHM}}l_{\text{SHM}})/(w_{\text{SHM}}t_{\text{SHM}}) 
\end{align*}

\noindent
Here, $C_g$, $C_{{\text{FE}}_{\text{in}}}$, $R_{on}$, and $R_{\text{SHM}}$ 
refer to the transistor gate capacitance, capacitance of the input FE 
capacitor, transistor on--resistance, and resistance of the SHM,
respectively. The length, width, and thickness of the SHM, are
respectively, given by
$l$, $w$, and $h$, each with subscript SHM. 

\subsection{Layout of CoMET--based majority gate}
The layout of MAJ3 corresponding to the schematic shown in Fig.~\ref{fig:majority} 
for a chosen value of $F$, is shown in Fig.~\ref{fig:layout}.
We draw the layout according to the design rules for $F$ as
described in detail in~\cite{BeyondCMOS}. The reset transistor for
each input FE capacitor $i$,  $T_{RST,i}$, with $i \in {1, 2, 3}$, and
the reset transistor for the output FE capacitor, $T_{RST,o}$, are local 
to the majority gate as shown in the layout. The transistor required to
send a charge current through the SHM, $T_{PROP}$, is shared globally by
multiple gates. The dual-rail inverter is local to the majority gate and
transfers the information to the next stage. The area of the MAJ3 gate
is $29F\times16F$ nm$^2$.      

%% file: sections/results.tex
\section{Results and Discussion}
\label{sec:results}

\begin{figure}[h]
\centering
	\includegraphics[width=6cm]{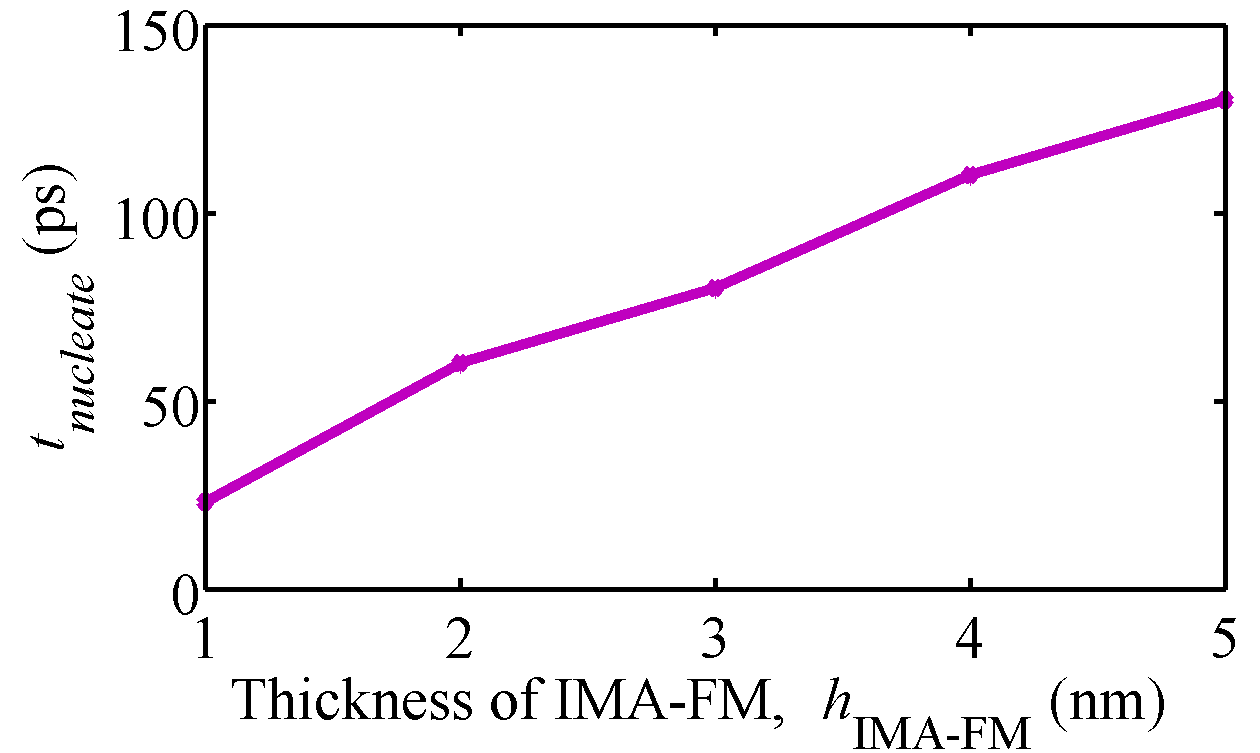}
	\caption{Nucleation delay, $t_{nucleate}$, of the CoMET device for
		$F=\SI{15}{nm}$, as a
		function of the IMA--FM thickness, $h_{\text{IMA--FM}}$. The PMA--FM material 
 		parameters used in the OOMMF simulation are:
		$M_{\text{S,PMA--FM}}= \SI{0.3e6}{A/m}$, $K_{\text{U,PMA--FM}} =
		\SI{0.5e6}{J/m^3}$, 
 		$A = \SI{10}{pJ/m}$ and $\alpha = 0.05$ (similar trends are seen for other
		parameter choices).} 
	\vspace{-8pt}
	\label{fig:imafm_thickness}
\end{figure} 

The delay of the device is a function of the dimensions of IMA--FM
and PMA--FM material parameters, specifically $M_{\text{S,PMA--FM}}$, $K_{\text{U,PMA--FM}}$,
$A$, and $\alpha$.  We explore the design space consisting of the combination
of these parameters and analyze their impact on device performance. We
demonstrate the results of the design space exploration for
$F=\SI{15}{nm}$ and show two sample design points for $F$ set to
\SI{15}{nm} and \SI{7}{nm}. 

\begin{figure}[h]
\centering
	  \subfigure[]{
          \includegraphics[width=6.5cm]{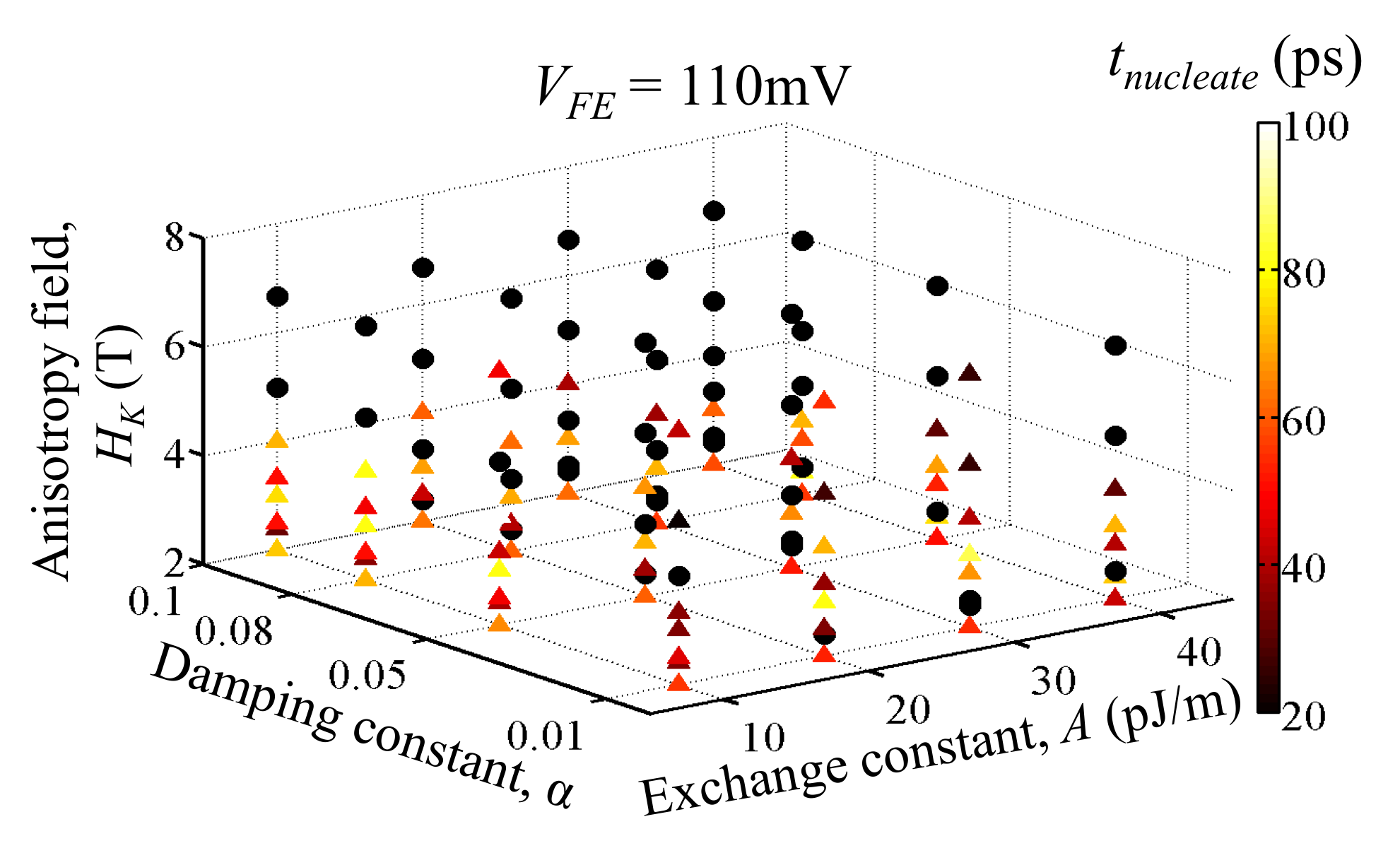}
     }
     \subfigure[]{
          \includegraphics[width=6.5cm]{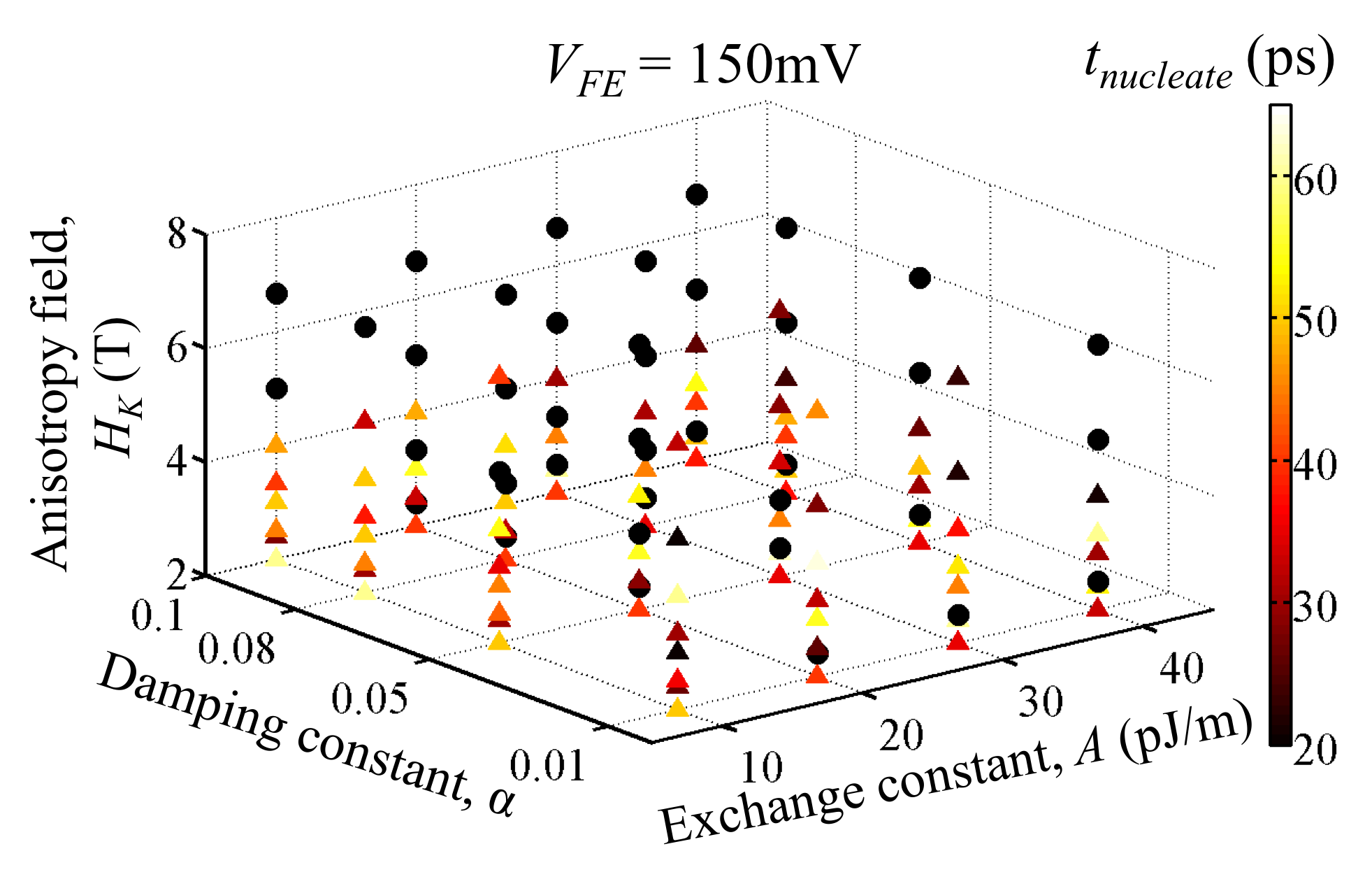}
     }
     \caption{Nucleation delay, $t_{nucleate}$, of the CoMET device for
$F=\SI{15}{nm}$ 
as a function of (a) material parameters for $V_{FE} = \SI{110}{mV}$ and (c)
material parameters for $V_{FE} = \SI{150}{mV}$. The triangles indicate successful 
 nucleation and while the circles indicate unsuccessful nucleation.}
     \label{fig:nuc}
\vspace{-8pt}
\end{figure}

\begin{figure*}[h]
     \centering
      \subfigure[]{
          \includegraphics[width=0.3\textwidth]{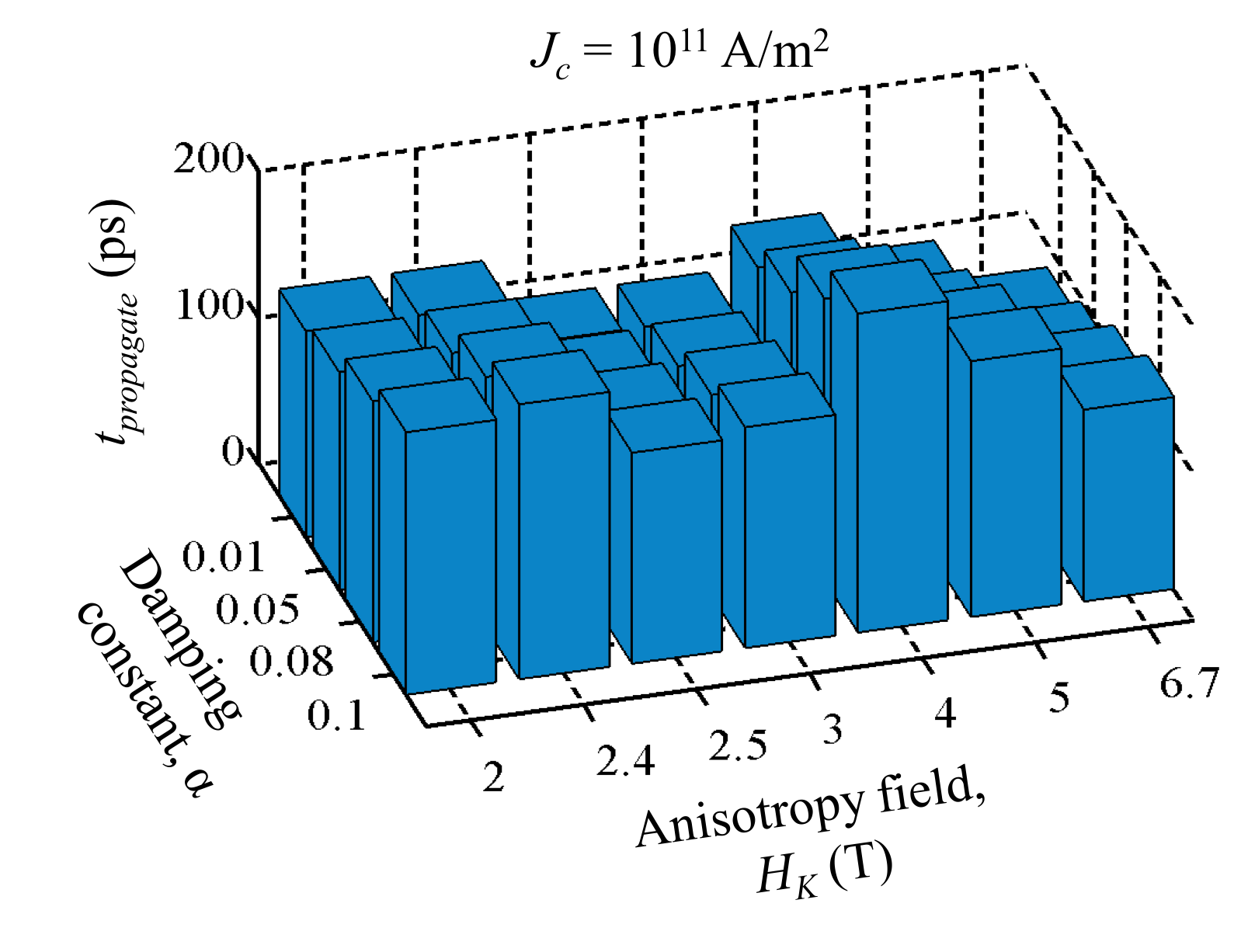}
     }
     \subfigure[]{
         \includegraphics[width=0.3\textwidth]{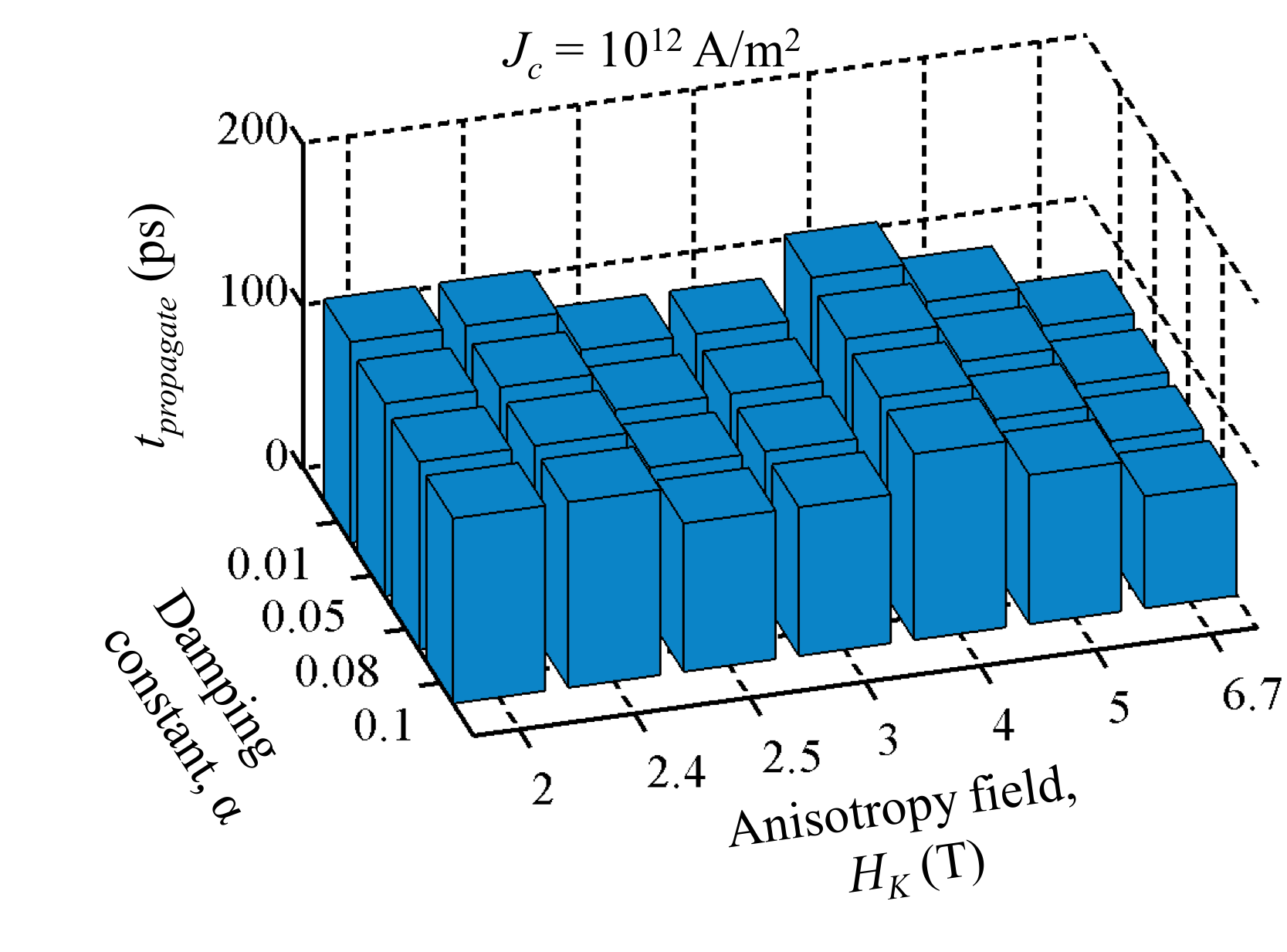}
     }
	\subfigure[]{
		\includegraphics[width=0.33\textwidth]{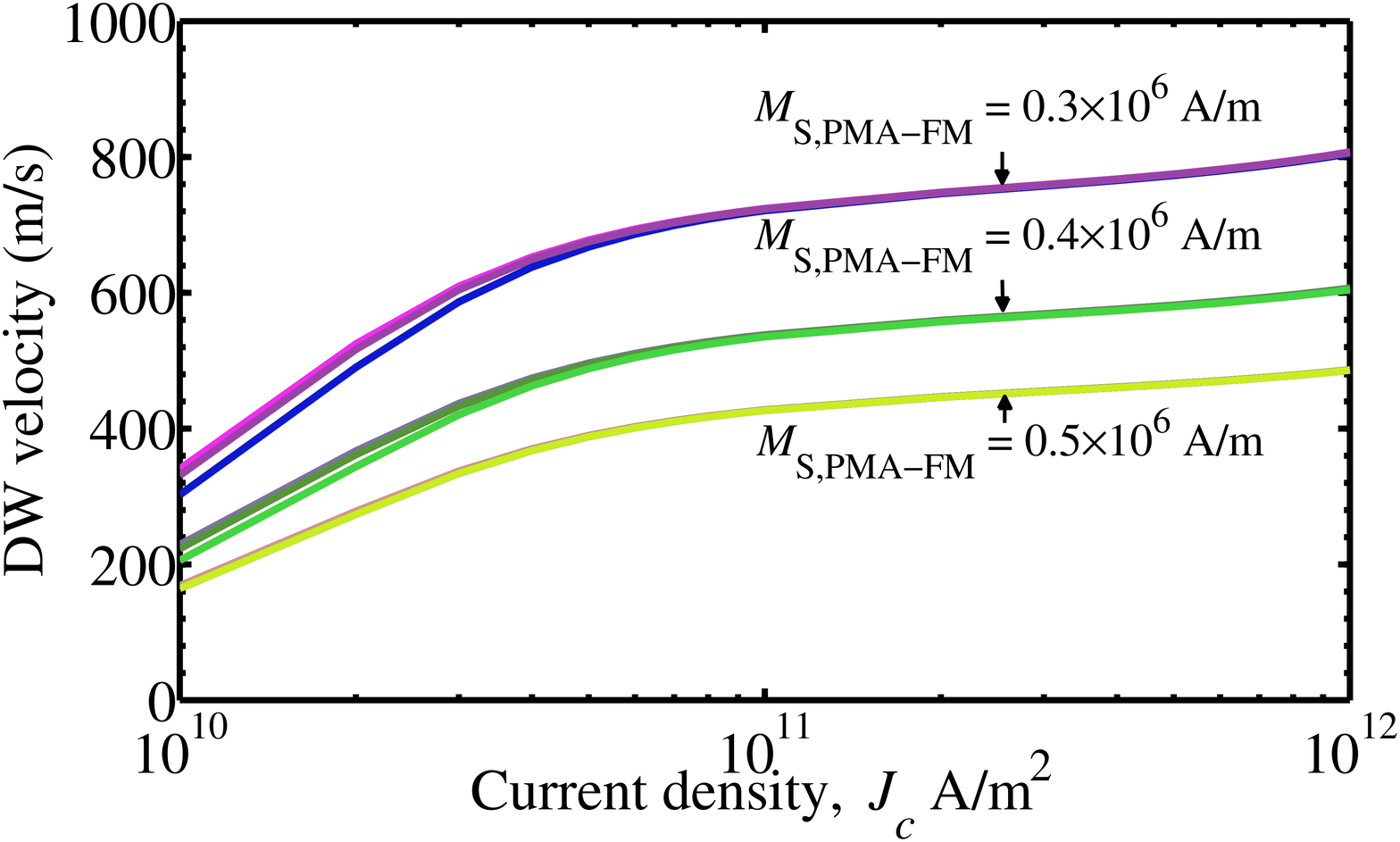}
	}
     \caption{DW propagation delay, $t_{propagate}$, of the CoMET
device for $F=\SI{15}{nm}$ as a function of the material 
     parameters for $A =\SI{10}{pJ/m}$ with (a) $J_c =\SI{e11}{A/m^2}$, (b)
$J_c =\SI{e12}{A/m^2}$ and (c)~DW velocity as a function of the current 
density, $J_c$, for all of the design points shown in (a) and (b). 
Note that points on the x and y axes of the bar chart in (a) and (b) 
are not equally spaced.}
	\vspace{-8pt}
     \label{fig:prop}
\end{figure*}

\subsection{Choice of material parameters}
The simulation parameters and their values used in this work are listed 
in Table~\ref{tbl:parameters}. The parameter space is chosen to reflect realistic values: 
the choice of $A \in \{\SI{10}{pJ/m}, \SI{20}{pJ/m}, \SI{30}{pJ/m},
\SI{40}{pJ/m}\}$ is chosen to
reflect the typical exchange constant of existing and exploratory
ferromagnetic materials. Lowering $A$ further would make the Curie
temperature too low~\cite{ohandley}. 
The choice of $M_{\text{S,PMA--FM}} \in \{\SI{0.3e6}{A/m}, 
\SI{0.4e6}{A/m}, \SI{0.5e6}{A/m}\}$ and
$K_{\text{U,PMA--FM}} \in \{\SI{0.5e6}{J/m^3},
\SI{0.6e6}{J/m^3}, 
\SI{1e6}{J/m^3}\}$ allow the mapping of PMA--FM materials to existing materials. 
The choice of $\alpha \in \{0.01, 0.05, 0.08, 0.1\}$ is free of any constraint to material mapping 
as it can be modified by adequately doping the PMA--FM~\cite{doping1,doping2}.
The saturation magnetization of the IMA--FM, $M_{\text{S,IMA--FM}}$ is set to
\SI{1e6}{A/m}. The value of $A$ and $\alpha$
for the IMA--FM is set to the same value as that of PMA--FM. 

\subsection{DW nucleation}
We estimate $t_{nucleate}$ in OOMMF when the DW nucleates beneath the
IMA--FM as shown in the snapshots in Fig.~\ref{fig:OOMMF}(a). We first 
relax the composite structure in OOMMF for \SI{200}{ps} before 
applying the effective ME field as a Zeeman field. This time period
allows the PMA--FM to reach an equilibrium state before the DW is
nucleated. In a typical circuit,
this state could be achieved by the PMA--FM in the time interval between 
successive switching activity. At the end of $\SI{200}{ps}$, denoted in
the figure as $t_1 = \SI{0}{ps}$, the magnetization of the 
PMA--FM rests at an angle to the easy axis owing to the strong exchange
coupling with the IMA--FM. After applying a Zeeman field,
the DW nucleates in PMA--FM at $2F$ after a delay of \SI{44}{ps}.

We compare the voltages required to nucleate the DW in the PMA--FM at
approximately the same $t_{nucleate}$, in the
absence of the IMA--FM on top of the PMA--FM to provide the initial
angle. The procedure to calculate $t_{nucleate}$ is identical to the
experiment in 
Fig.~\ref{fig:OOMMF}(a). We perform this analysis for two cases: 
(i) when the applied Zeeman field acts on a region $2F\times1F\times1$nm 
corresponding to the scenario shown in Fig.~\ref{fig:OOMMF}(b). The DW 
nucleates at $t_{nucleate} = \SI{44}{ps}$ at $2F$. However,
$V_{FE}$ required to generate the DW is now \SI{350}{mV}. After
relaxing the magnetization for $\SI{200}{ps}$, an absence of
IMA--FM translates to a very low initial angle at $t_1 =
\SI{0}{ps}$ which
necessitates a stronger effective ME field, $H_{ME}$, and therefore a higher $V_{FE}$ to
nucleate the DW for a given $t_{nucleate}$. (ii) The absence of IMA--FM 
allows us to further compact the CoMET device such that the FE capacitor 
dimensions are the minimum possible at a chosen value of $F$. This
corresponds to the dimensions, $1F\times1F\times5$nm (as opposed to those 
shown in Fig.~\ref{fig:majority}(c)), the region from the left
end of PMA--FM on which $H_{ME}$ acts. We find that the voltage
required to nucleate the DW at $1F$, as shown in
Fig.~\ref{fig:OOMMF}(c), 
is close to \SI{1}{V}. From these two experiments, we conclude that the composite 
structure facilitates a fast and energy-efficient DW nucleation.   

The nucleation of DW in the PMA--FM is not only a function of PMA--FM
material parameters, but also depends on the material dimensions of
the IMA--FM. As stated in Section~\ref{sec:modeling}, the aspect ratio of the IMA--FM is
set to 2:1 to obtain the shape anisotropy necessary for the coupling
with PMA--FM. We then explore the dependence of $t_{nucleate}$ on the thickness of
IMA--FM, $h_{\text{IMA--FM}}$ and plot the results in
Fig.~\ref{fig:imafm_thickness}. As $h_{\text{IMA--FM}}$ increases, 
it becomes harder to switch the PMA--FM due to strong exchange coupling 
between IMA--FM and PMA--FM, increasing $t_{nucleate}$. 
We therefore select $h_{\text{IMA--FM}} = \SI{1}{nm}$. 

The impact of material parameters of PMA--FM 
on $t_{nucleate}$ is shown in Fig.~\ref{fig:nuc}(a) and
Fig.~\ref{fig:nuc}(b) 
for $V_{FE}=\SI{110}{mV}$ and $V_{FE}=\SI{150}{mV}$, respectively.
It is seen that (a)~a larger $V_{FE}$ reduces $t_{nucleate}$, and this
can be shown to be consistent with the DW nucleation
Equations~(\ref{eq:LKh}--\ref{eq:LLG}). A larger $V_{FE}$ corresponds to
a larger $E_{FE}$ across FE$_{\text{in}}$, which in turn creates a
larger $H_{ME}$ to nucleate the DW faster. (b)~Lower values of $H_{K}$ are 
more conducive to nucleation; a lower anisotropy field makes it easier
for $H_{ME}$ to switch the magnetization between the two easy axes and 
(c)~low values of $A$ reduce $t_{nucleate}$ owing to the weaker exchange coupling with the
neighboring magnetic domains of the PMA--FM. We note that for
$A>\SI{10}{pJ/m}$, the number of design points at which the nucleation
does not occur increases. Therefore we pick the lowest
value of $A=\SI{10}{pJ/m}$. This choice does not restrict the design search 
space for DW propagation as $t_{propagate}$ is primarily dictated by the 
choice of $M_{\text{S,PMA--FM}}$. 

\subsection{DW propagation}
With this choice, we show $t_{propagate}$ for $J_c={10^{11}}$ A/m$^2$ and $10^{12}$ A/m$^2$
in Figs.~\ref{fig:prop}(a) and (b), respectively. Increasing $J_c$
increases the torque from SHE as seen from the expressions for
$H_{SHE}$ in Equation~\ref{eq:fields}. This can be seen
from Fig.~\ref{fig:prop}(c) where increasing $J_c$ increases the DW
velocity, thereby reducing $t_{propagate}$.  These curves lie in three clusters, and
show the dominance of $M_{\text{S,PMA--FM}}$ over other parameters. 

This can also be seen from Figs.~\ref{fig:prop}(a) and (b) where the lowest 
$t_{propagate}$ bars ($H_K = \SI{6.7}{T}$) correspond to
$M_{\text{S,PMA--FM}}=\SI{0.3e6}{A/m}$. This is
consistent with the Equation~\ref{eq:fields}: a lower $M_{\text{S,PMA--FM}}$
implies higher $H_{SHE}$ and $H_{DMI}$, and therefore higher DW velocity.  

 \begin{figure}[h]
    \centering 
    \subfigure[] {
        \includegraphics[width=7cm]{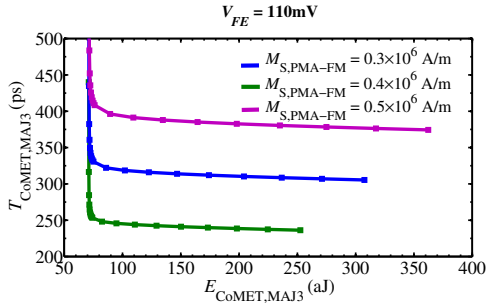}
    }
    \subfigure[] {
        \includegraphics[width=7cm]{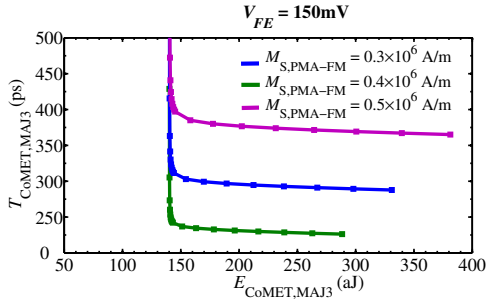}
    }
     \caption{Energy vs. delay of the CoMET--based MAJ3 gate for three design points, corresponding to the 
    $M_{\text{S,PMA--FM}}$ values for the clusters in
    Fig.~\ref{fig:prop}(c) for (a) $V_{FE}$. Other parameter values: 
$\alpha = 0.01$, $A = \SI{10}{pJ/m}$, $K_{\text{U,PMA--FM}} =
\SI{0.5e6}{J/m^3}$.}
    \vspace{-8pt}
	\label{fig:edp}
\end{figure}

\subsection{Performance evaluation}
For the three $M_{\text{S,PMA}}$ corresponding to each of the three
clusters in Fig.~\ref{fig:prop}(c), we plot $T_{\text{CoMET}}$ vs.
$E_{\text{CoMET}}$ for MAJ3 in Fig.~\ref{fig:edp} for the two values of
$V_{FE}$. The dual--rail inverter delay, $t_{qtransfer}$, is calculated using the PTM
technology models~\cite{PTM}. For a chosen $M_{\text{S,PMA--FM}}$ and
$V_{FE}$, the energy-delay data points are obtained by increasing $J_c$
from $\SI{e10}{A/m^2}$ to $\SI{e12}{A/m^2}$ in discrete steps. The main
observations from Fig.~\ref{fig:edp} are as follows:
\begin{itemize}  
\item
Increasing $V_{FE}$ is seen to reduce $T_{\text{CoMET}}$ by reducing $t_{nucleate}$, at the
expense of a larger $E_{\text{CoMET}}$.  
\item
A higher $J_c$ corresponds to lower $T_{\text{CoMET}}$, but $E_{\text{CoMET}}$ 
is only marginally higher since it is primarily dominated by the
transistor energy. 
\item 
Initially when $J_c$ increases, $T_{\text{CoMET}}$ reduces at the same rate 
as $J_c$, thus keeping the energy approximately constant. After a certain
point, increasing $J_c$ only gives marginal improvements in delay. This 
result is consistent with Fig.~\ref{fig:prop}(c); as $J_c$ increases
from $10^{10}$ A/m$^2$ to $10^{12}$ A/m$^2$, DW velocity increases sharply
initially but only increases gradually later. 
\item
A robust design point can be chosen such that $T_{\text{CoMET}}$ is less variable 
with material parameters. This corresponds to the right portion of each
curve where the delay only improves marginally with increase in $J_c$.
\end{itemize}

\begin{table}[]
\footnotesize
\centering
\begin{center}
\begin{tabular}{|p{0.65cm}|p{1.2cm}|p{1.3cm}|p{1.4cm}|p{1.3cm}|}
\hline
$V_{FE}$ (mV) & $t_{nucleate}$ (ps) & $t_{propagate}$ (ps) & $t_{qtransfer}$ (ps) & $T_{\text{CoMET}}$ (ps) \\ \hline
110           & 35/35               & 77.4/38.7            & 8.8/8.8                & 242.4/165.5            \\ \hline
150           & 30/30               & 77.4/38.7            & 8.2/8.2                & 231.2/153.8            \\ \hline
\end{tabular}
\\[2pt]
\begin{tabular}{|p{0.5cm}|p{0.8cm}|p{0.9cm}|p{1cm}|p{1cm}|p{1.2cm}|}
\hline
$V_{FE}$ (mV) & $E_{FE}$ (aJ) & $E_{TX}$ (aJ) & $E_{Joule}$ (aJ) & $E_{leakage}$ (aJ) & $E_{\text{CoMET}}$ (aJ) \\ \hline
110           & 2.4/0.8       & 40.8/24.2     & 19.8/1.6         & 16.3/16.3          & 158.6/85.8             \\ \hline 
150           & 4.4/1.5       & 42.0/30.6     & 25.5/1.5         & 22.8/22.8          & 189.4/112.8            \\ \hline
\end{tabular}
\\[2pt]
(A)
\\[2pt]
\bigskip
\begin{tabular}{|p{0.65cm}|p{1.2cm}|p{1.3cm}|p{1.4cm}|p{1.3cm}|}
\hline
$V_{FE}$ (mV) & $t_{nucleate}$ (ps) & $t_{propagate}$ (ps) & $t_{qtransfer}$ (ps)    & $T_{\text{CoMET}}$ (ps) \\ \hline
110           & 30/30               & 36.2/18.1            & 7.9/7.9                 & 148.2/112.0            \\ \hline
150           & 25/25               & 36.2/18.1            & 6.2/6.2                 & 134.8/98.6             \\ \hline
\end{tabular}
\\[2pt]
\begin{tabular}{|p{0.5cm}|p{0.8cm}|p{0.9cm}|p{1cm}|p{1cm}|p{1.2cm}|}
\hline
$V_{FE}$ (mV) & $E_{FE}$ (aJ) & $E_{TX}$ (aJ) & $E_{Joule}$ (aJ) & $E_{leakage}$ (aJ) & $E_{\text{CoMET}}$ (aJ) \\ \hline
110           & 0.5/0.1       & 16.8/12.0     & 1.8/0.1          & 13.7/13.7          & 65.6/51.8              \\ \hline
150           & 0.9/0.3       & 21.4/15.3     & 1.8/0.1          & 18.5/18.5          & 85.2/68.4              \\ \hline
\end{tabular}
\\[2pt]
(B)
\end{center}
\caption{Delay and energy of CoMET--based MAJ3/INV gate for
(A)~$F=\SI{15}{nm}$
and (B)~$F=\SI{7}{nm}$ for the design
point corresponding to parameters, $M_{\text{S,PMA--FM}}=\SI{0.3e6}{A/m}$, 
$K_{\text{U,PMA--FM}}=\SI{0.5e6}{J/m^3}$, $J_c=\SI{5e11}{A/m^2}$,
$\alpha = 0.01$, and
$A=\SI{10}{pJ/m}$.}
\label{tbl:tbl_edp_15nm}
\end{table}

\ignore{
\begin{figure}[h]
\centering
\subfigure[]{
	\includegraphics[width=7.5cm]{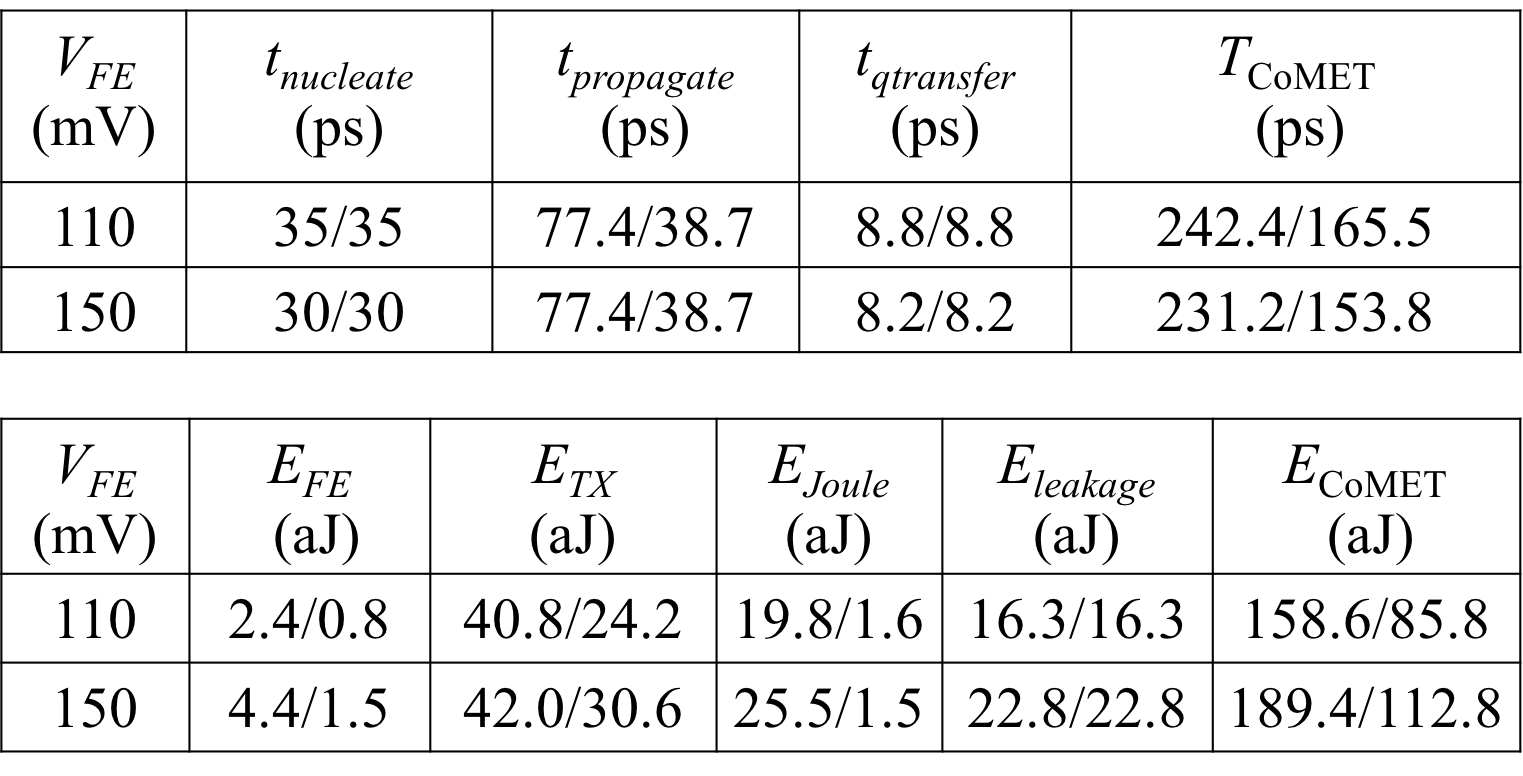}
}
\subfigure[]{
	\includegraphics[width=7.5cm]{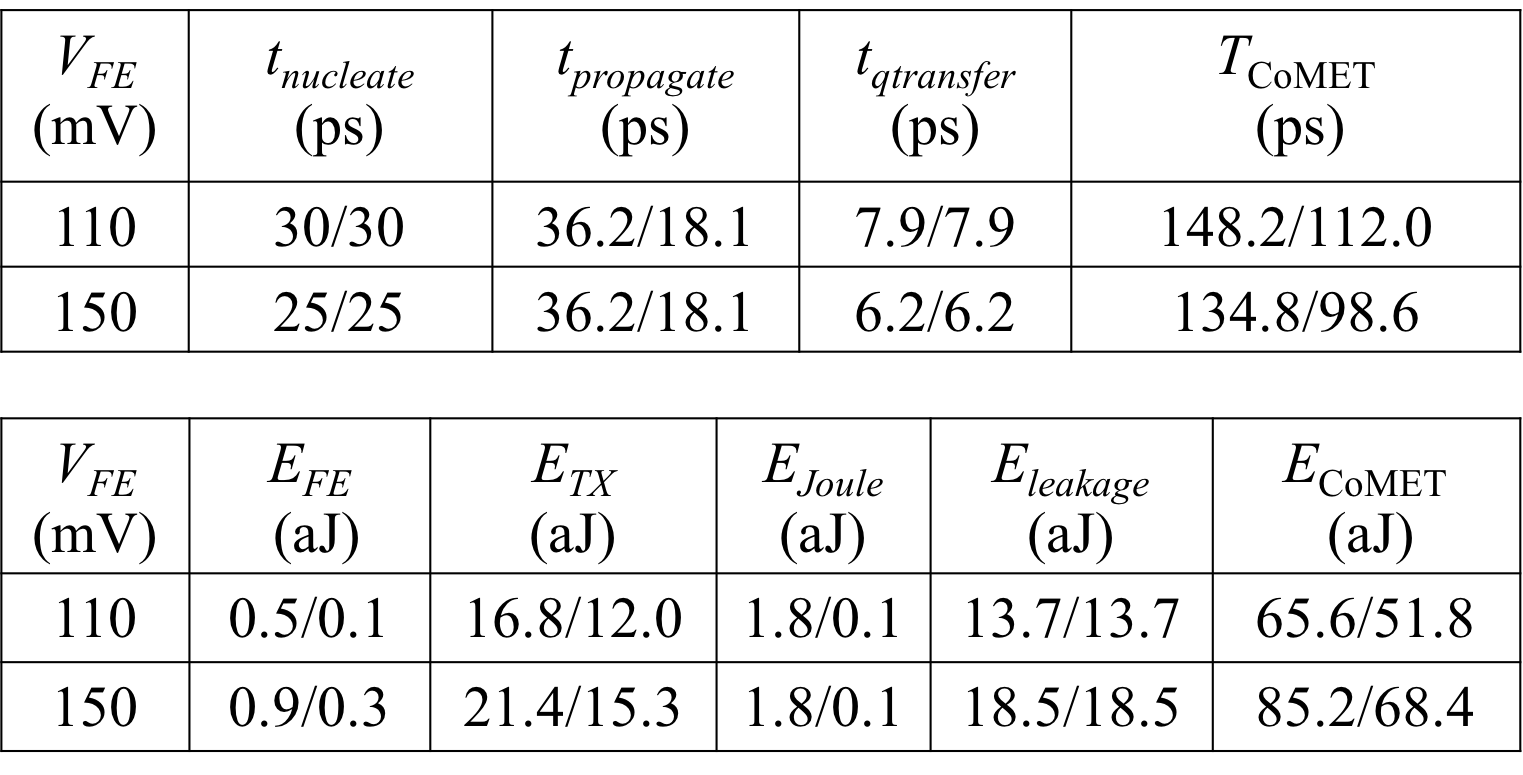}
}
\caption{Delay and energy of CoMET--based MAJ3/INV gate for
(a)~$F=\SI{15}{nm}$
and (b)~$F=\SI{7}{nm}$ for the design
point corresponding to parameters, $M_{\text{S,PMA--FM}}=\SI{0.3e6}{A/m}$, 
$K_{\text{U,PMA--FM}}=\SI{0.5e6}{J/m^3}$, $J_c=\SI{5e11}{A/m^2}$,
$\alpha = 0.01$, and
$A=\SI{10}{pJ/m}$.}
\vspace{-8pt}
\label{fig:tbl_edp_15nm}
\end{figure}
}

The best ($T_{\text{CoMET}}$, $E_{\text{CoMET}}$) for each $V_{FE}$ for
MAJ3/INV for $F=\SI{15}{nm}$ and $F=\SI{7}{nm}$ are shown in
Table ~\ref{tbl:tbl_edp_15nm}(A) and (B), respectively. It can be seen
that $t_{propagate}$ dominates $T_{\text{CoMET}}$ while
$E_{\text{CoMET}}$ is dominated by energy associated with turning the
transistors on and the corresponding leakage. The delay and energy
obtained using the CMOS technology given respectively by ($T_{\text{CMOS}}$,
$E_{\text{CMOS}}$) for an inverter is ($\SI{1.8}{ps}, \SI{38.7}{aJ}$) at $F=\SI{15}{nm}$ and 
($\SI{1.6}{ps}, \SI{19.8}{aJ}$) at $F=\SI{7}{nm}$. For
CMOS-based MAJ3 gate, the performance numbers are (14.8ps, 704.2aJ) at
$F=\SI{15}{nm}$ and ($\SI{11.4}{ps}, \SI{361.6}{aJ}$) at $F=\SI{7}{nm}$. 
The CMOS performance numbers were obtained using the PTM technology models{~\cite{PTM}} 
at nominal supply voltages of $\SI{0.85}{V}$ for $F=\SI{15}{nm}$ and 
$\SI{0.7}{V}$ for $F=\SI{7}{nm}$. Thus we see that a MAJ3 gate can be 
implemented more energy-efficiently with CoMET than with CMOS.

At these design points, $M_{\text{S,PMA--FM}}$, $K_{\text{U,PMA--FM}}$, and $A$
can be mapped to MnGa--based Heusler alloy~\cite{MnGa,Fiebig2005}.  The damping
constant, $\alpha = 0.01$ can be engineered by choosing a new composition of PMA--FM.  For the FE
layer, BiFeO$_3$ (BFO) can be used~\cite{IntelME}, while the SHM could be
$\beta$-W, Pt, $\beta$-Ta~\cite{sheTa,shePt,sheW} or some new materials under investigation.

%% file: sections/conclusion.tex
\section{Conclusion}
\label{sec:conclusion}

A novel spintronic logic device based on magnetoelectric effect and fast current--driven 
domain wall propagation has been proposed. We have shown that the composite input structure of a FM with 
IMA placed in contact with a PMA--FM allows circuit operation at low voltages of \SI{110}{mV} and \SI{150}mV. 
A novel circuit structure comprising 
a dual--rail inverter structure for efficient logic cascading has also been introduced. The impact 
of the different material parameters on the performance of the device is then systematically 
explored. An optimized INV has a delay of \SI{98.6}{ps} with energy dissipation of \SI{68.4}{aJ} at 7nm, 
while a MAJ3 gate runs at \SI{134.8}{ps} and \SI{85.2}{aJ}.

%% file: sections/ack.tex
\section*{Acknowledgment}

This work was supported in part by C-SPIN, one of the six SRC STARnet Centers,
sponsored by MARCO and DARPA. The authors thank Dr. Angeline Klemm Smith and Dr. Chenyun Pan 
for their inputs.